\documentclass[journal=langd5,manuscript=article]{achemso}

\usepackage{achemso}
\setkeys{acs}{usetitle=true}
\SectionNumbersOn
\mciteErrorOnUnknownfalse

\usepackage[version=3]{mhchem} 
\usepackage{natbib}
\usepackage{graphicx}
\usepackage{epstopdf}
\usepackage{psfrag}
\usepackage{amsmath}
\usepackage{eucal}
\usepackage{booktabs}

\usepackage{amssymb}
\usepackage{txfonts}
\usepackage{dashrule}
\usepackage{pifont} 
\usepackage{wasysym} 
\author{Stefan Becker}
\affiliation[Laboratory of Engineering Thermodynamics]
{Laboratory of Engineering Thermodynamics, University of Kaiserslautern, Erwin-Schr\"odinger-Stra{\ss}e 44, 67663 Kaiserslautern, Germany}
\author{Herbert M. Urbassek}
\affiliation[Physics Department and Research Center OPTIMAS]
{Physics Department and Research Center OPTIMAS, University of Kaiserslautern, Erwin-Schr\"odinger-Stra{\ss}e 46, 67663 Kaiserslautern, Germany}
\author{Martin Horsch}
\email{martin.horsch@mv.uni-kl.de}
\affiliation[Laboratory of Engineering Thermodynamics]
{Laboratory of Engineering Thermodynamics, University of Kaiserslautern, Erwin-Schr\"odinger-Stra{\ss}e 44, 67663 Kaiserslautern, Germany}
\author{Hans Hasse}
\affiliation[Laboratory of Engineering Thermodynamics]
{Laboratory of Engineering Thermodynamics, University of Kaiserslautern, Erwin-Schr\"odinger-Stra{\ss}e 44, 67663 Kaiserslautern, Germany}

\title[Contact angle in LJ systems]
{Contact angle of sessile drops in Lennard--Jones systems}

\begin{document}
\begin{abstract}
  Molecular dynamics simulations are used for studying the contact angle of nanoscale sessile drops on a planar solid wall 
  in a system interacting via the truncated and shifted Lennard--Jones potential.
  The entire range between total wetting and dewetting is investigated by varying the solid--fluid dispersive interaction energy. 
  The temperature is varied between the triple point and the critical temperature.
  A correlation is obtained for the contact angle in dependence of the temperature and the dispersive interaction energy.
  Size effects are studied by varying the number of fluid particles at otherwise constant conditions, using up to 150000 particles.
  For particle numbers below 10000, a decrease of the contact angle is found.
  This is attributed to a dependence of the solid--liquid surface tension on the droplet size.
  A convergence to a constant contact angle is observed for larger system sizes. 
  The influence of the wall model is studied by varying the density of the wall.
  The effective solid--fluid dispersive interaction energy at a contact angle of $\theta = 90^\circ$ is found 
  to be independent of temperature and to decrease linearly with the solid density.
  A correlation is developed which describes the contact angle as a function of the dispersive interaction, the temperature and 
  the solid density.
  The density profile of the sessile drop and the surrounding vapor phase is described by a correlation combining a 
  sigmoidal function and an oscillation term.  
\end{abstract}


\section{Introduction \label{secIntro}}
Wetting of a solid phase by a liquid plays an important role in many processes.
The equilibrium wetting behavior is often classified by the contact angle $0^\circ \leq \theta \leq 180^\circ$ of a sessile drop.
The contact angle depends on the interaction between the particles, namely the fluid--fluid and the solid--fluid interactions.
These can be explicitly described with force fields and, hence, the force fields yield the contact angle.
While much work is available on force fields which describe the interaction in fluids 
\cite{Jorgensen1988, Jorgensen1996, Martin1998,Keasler2012, PANAGIOTOPOULOS1998,  POTOFF1999, MacKerell2000, Vrabec2001, Stoll2003, Vorholz2004, Moghaddam2004, Deublein2012, 
Merker2012},
solid--fluid interactions have been studied less systematically. 
In that field, mainly adsorption of simple fluids in nanopores is considered
\cite{Schapotschnikow2007a,Steele1974, Findenegg1975, Fischer1982, Bucior2009, Schapotschnikow2009, Sokolowski1990}
which enables fitting model parameters to adsorption isotherms. 
There are also reports on predicting the contact angle with force fields both for droplets 
\cite{Sikkenk1987, Sikkenk1988, Nijmeijer1989, Nijmeijer1990, Nijmeijer1992,  Tang1995, Blake1997, Werder2001, Werder2003, Ingebrigtsen2007, 
Grzelak2010a, Leroy2010,Rane2011}, and for fluid cylinders\cite{Weijs2011, Shahraz2012}.
However, they are restricted to few particular material combinations such as water on graphene.

The present work is devoted to studying the influence of the dispersive solid--fluid interaction on the contact angle in a model system 
by molecular dynamics (MD) simulations.
This model system consists of a single sessile drop on a planar wall.
The truncated and shifted Lennard--Jones (LJTS) potential \cite{Allen2009} is used for describing the fluid--fluid, 
solid--solid as well as the solid--fluid interactions, extending previous studies on interfacial properties of the LJTS fluid 
\cite{Horsch2010, Vrabec2006,Meel2008}. 
The solid--fluid interaction and the temperature are varied and a quantitative correlation describing their influence on the contact angle 
is presented. 
The density of the solid substrate affects the total potential of the solid--fluid interaction by the number 
of interaction sites located in a certain distance to a fluid particle \cite{Hamaker1937}.
This is examined in simulations with solids of varying densities.
A correlation is established for predicting the contact angle as a function of temperature, solid--fluid dispersive interaction, and solid density.
The findings are discussed in the context of the results from different studies on the wetting behavior of Lennard--Jones (LJ) fluids 
\cite{Nijmeijer1990, Nijmeijer1992, Tang1995, Ingebrigtsen2007, Bucior2009, Horsch2010, Grzelak2010a}.
Furthermore, an empirical correlation is presented that qualitatively describes the density profile 
of a sessile drop on a planar wall.

The system sizes accessible to MD simulation are getting closer to the smallest experimental settings, 
but systematic MD studies like the one carried out in the present study are still limited to nanoscale scenarios.
When dealing with wetting phenomena on the nanoscale, one has to consider effects such as the line tension \cite{Pethica1977} 
or a decrease in the liquid--vapor surface tension due to the strong curvature of the interface \cite{Tolman1949}.
In the present study, a brute force approach is used to deal with this: 
The system size is increased until no dependence of the contact angle on the size is observed.
The number of fluid particles finally used is 15000, 
which is large enough to ensure that a further increase would not lead to significantly different results.

The paper is organized as follows:
In section~\ref{secModMethod}, the molecular model and the simulation method are described.
The results regarding the size effects, the density profile, the contact angle and the influence of the wall density on the contact angle 
are presented in section~\ref{secSimRes} and discussed in section~\ref{secDiscussion}.
Conclusions are drawn in section~\ref{secConclusions}.
Additional information is presented in the supporting information.

\section{Model and Simulation Method \label{secModMethod}}

\subsection{Molecular Model}

Like the original LJ potential 
$u^\mathrm{LJ}(r_{ij}) = 4 ~\epsilon [\left(\sigma/r_{ij} \right)^{12} - \left( \sigma/r_{ij} \right)^{6} ]$
the LJTS potential \cite{Allen2009} 
\begin{equation}
	u^\mathrm{LJTS}(r_{ij}) = \left\{ \begin{array}{lc}	u^\mathrm{LJ}(r_{ij}) - u^\mathrm{LJ}(r_c), & ~~r_{ij} < r_c \\
								0,  & ~~r_{ij}  \geq r_c
					    \end{array}
				    \right.
\label{u_LJTS}
\end{equation}
with a cutoff radius of $r_c = 2.5 ~\sigma$ can accurately reproduce the thermophysical properties of simple nonpolar fluids,
especially noble gases and methane \cite{Vrabec2006}.
It is used in the present study to describe all the three interaction types, i.e. fluid--fluid, solid--solid and solid--fluid.

The accurate description of solids usually requires the use of multibody potentials that are computationally more expensive \cite{Engin2008}.
The present study, however, is not concerned with the properties of a solid phase 
but rather with the influence of the solid--fluid interaction on the fluid, if solely dispersive and repulsive interactions are present.
The wall is represented here by particles arranged in a face--centered cubic (fcc) lattice with the (100) surface exposed to the fluid.
To maintain the wall in the solid state, the LJ energy parameter of the solid ($s$) is related to that of the fluid ($f$) by 
$\epsilon_s = 100 ~\epsilon_f$ which essentially yields a static lattice.
With the size parameter of the solid $\sigma_s$, 
the lattice constant of the solid phase is $a = 1.55~\sigma_s$ and the particle density is  $\rho_{s} = 1.07 ~\sigma_s^{-3}$. 
It may be noted that the present choice of the cutoff radius, i.e. $r_c = 2.5~\sigma_f$, yields practically the same lattice constant 
as would have been obtained for $r_c \xrightarrow{} \infty$, i.e. for the full LJ potential.
Unless stated otherwise, the size parameters of the solid and the fluid are the same in the present study, i.e. $\sigma_s = \sigma_f$.
For a set of simulations in which the influence of the solid density is studied, the LJ size parameter of the solid $\sigma_s$ is varied.
By scaling down $\sigma_s$, the lattice constant of the solid is decreased and, hence, the density is increased: 
For the size parameters $\sigma_s = 0.80~\sigma_f$ and $\sigma_s = 0.646~\sigma_f$, the solid density is $\rho_s = 2.10~\sigma_f^{-3}$ and 
$\rho_s = 4.02~\sigma_f^{-3}$, respectively.

The dispersive and the repulsive interaction between the solid and the fluid phase is also described by the LJTS potential.
The LJ size parameter of the unlike interaction between solid and fluid particles ($sf$) is chosen to be $\sigma_{sf} = \sigma_{f}$.
Note that $\sigma_{sf} = \sigma_f$ even holds in the cases where the size parameter $\sigma_s$ is varied.
The LJ energy parameter of the solid--fluid interaction is scaled by 
\begin{equation}
\epsilon_{sf} = \zeta \epsilon_f. 
\label{eq:unlikeInteraction}
\end{equation}
$\zeta$ is called reduced solid--fluid interaction energy. Its influence on the contact angle is studied systematically.

\subsection{Simulation Method}
MD simulations in the canonical ensemble are carried out with the program \textit{ls1 MarDyn} \cite{Buchholz2011124} to obtain the 
contact angle dependence on the temperature and the reduced solid--fluid interaction energy.
Velocity Scaling was applied for temperature control.
The parallelization is accomplished by spatial domain decomposition based on a linked cell data structure.
Newton's equations of motion are integrated via the Verlet leapfrog algorithm with a  time step of 
$5 \cdot 10^{-4}~\epsilon_f^{- 1/2}~m^{1/2}~\sigma_f$.

A sessile drop on a planar solid wall is simulated (see shanpshot in Figure~S.1 in the supporting information). 
The system contains a single drop, i.e. there is no drop on the opposite side of the wall.
This corresponds to the stable configuration in  the entire regime of partial wetting ($0^\circ < \theta < 180^\circ$) \cite{Sikkenk1988}.
The wall is located in the $(x,z)$--plane, and $y$ represents the distance from the 
plane in which the centers of the uppermost wall particles lie. 
Periodic boundary conditions are applied in all directions, leaving a channel for the fluid between the wall and its
periodic image.
The size of the simulation box is adapted such that for small contact angles the fluid has sufficient space in lateral dimensions.
On the other hand, when large contact angles are expected, the spacing of the channel is chosen sufficiently large 
to avoid a perturbation of the droplet by the periodic image of the wall.
The height of the channel exceeds $30 ~\sigma_f$ in all cases, which avoids artifacts due to finite size effects \cite{Oleinikova2006} related to the
channel height.
The number of wall particles varies according to the box dimensions.
The wall thickness of two and a half crystal unit cells exceeds the cutoff radius of the fluid and thus avoids periodic artefacts,  
which could be caused by an interaction of fluid particles on different sides of the wall.
The initial configuration of the system consists of a cuboid with liquid density \cite{Vrabec2006} 
and a surrounding vapor phase, both in contact with the wall.
The number of fluid molecules is 15000 except for a series of simulations conducted to investigate size effects, in which this number is varied.
The equilibration time is at least 2.5 million time steps, followed by 3.5 million time steps of production.

\begin{figure}[ht!]
\includegraphics[height = 8.333cm]{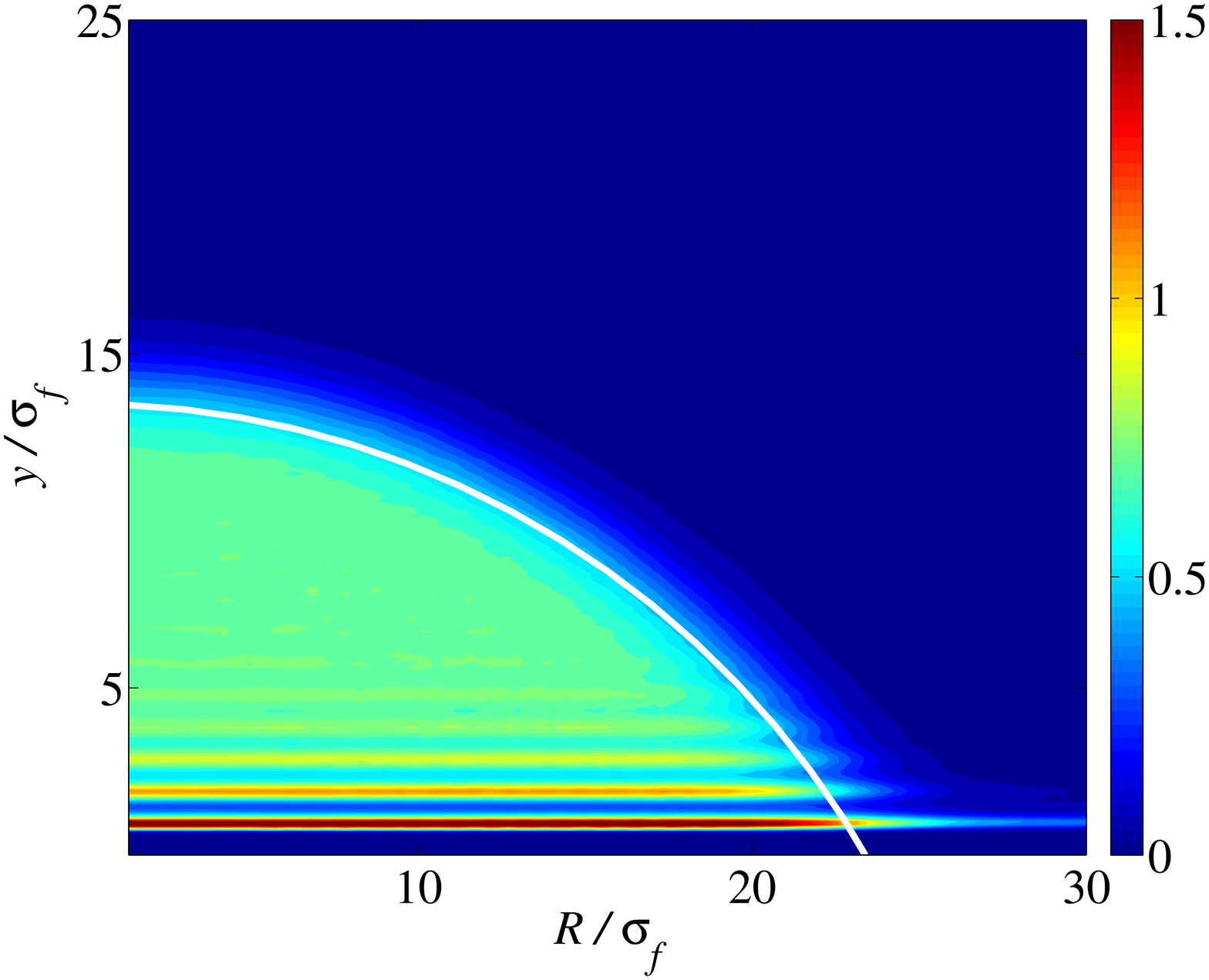}
\includegraphics[height = 8.333cm]{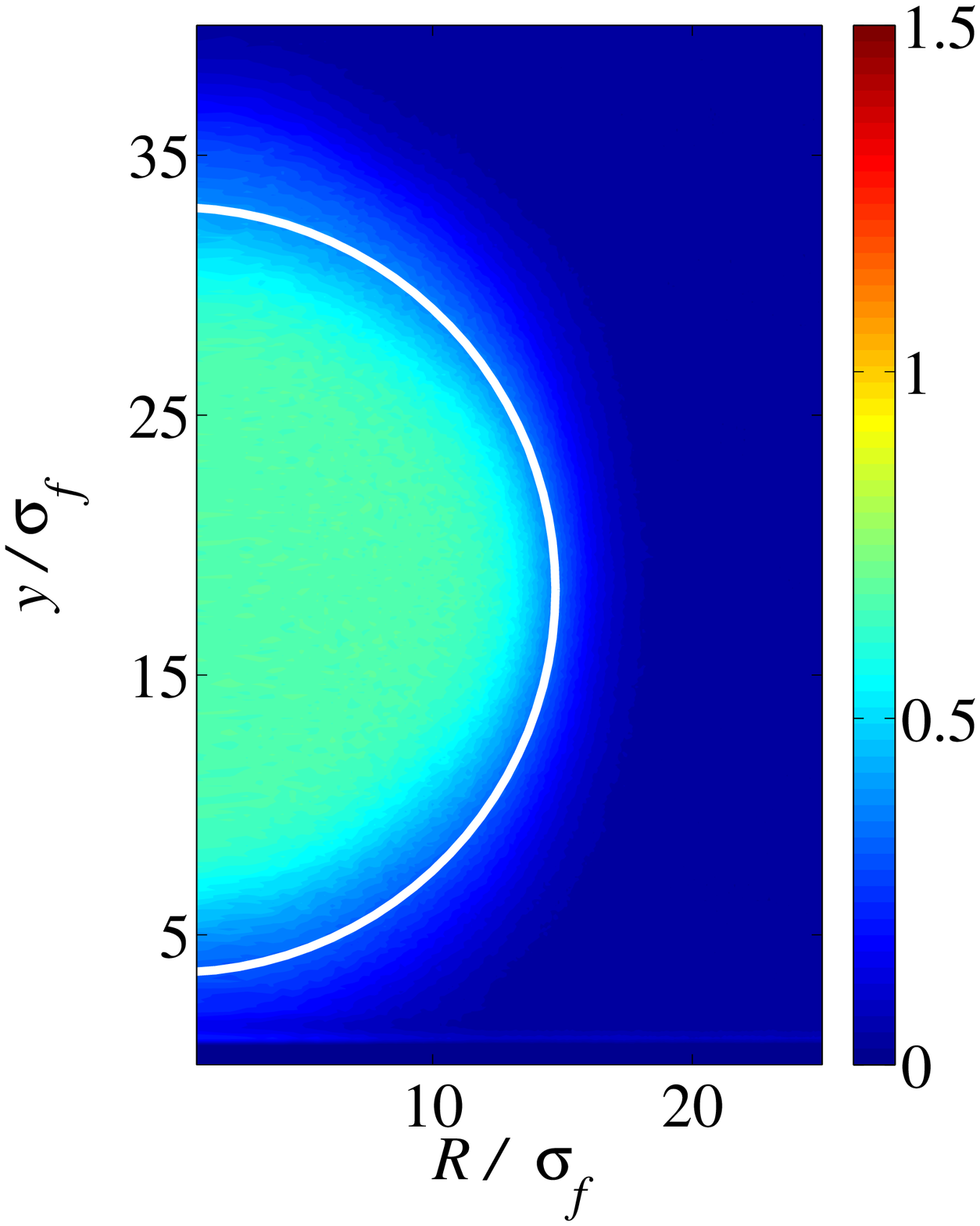}
\caption{Density profiles of liquid droplets. 
	 The color indicates the density in units of $\sigma_f^{-3}$.
	 The white circle is fitted to the interface at the positions with $\rho = (\rho' + \rho'')/2$ 
	 and employed to determine the contact angle by extrapolating to $y = 0$.
	 Left: The reduced solid--fluid interaction energy is $\zeta = 0.65$ 
	 and the temperature $T = 0.8~\epsilon_f/ k$. 
	 An adsorbed fluid phase can be observed next to the droplet.
	 Right: Density profile of a droplet under conditions of total dewetting. 
	 The simulation parameters are $\zeta = 0.25$ and $T = 0.9 ~\epsilon_f/ k$.}
\label{fig:SampledDensProfile2D}
\end{figure}

The sampling of the density profile $\rho(R,y)$ during the simulation is accomplished via binning in a cylindrical coordinate system, 
in terms of the distance from the wall $y$ and the distance from the symmetry axis of the droplet $R$.
In the vicinity of the solid wall, the fluid is affected by strong ordering effects. 
By choosing a bin size of $0.1~\sigma_f$ in the direction normal to the wall, these effects are monitored.
The liquid-vapor interface is defined by the arithmetic mean density $\left( \rho' + \rho''\right)/2$ where $\rho'$ and $\rho''$ are the 
saturated bulk densities of the LJTS fluid known from previous studies \cite{Vrabec2006}.
As will be shown in section \ref{secSimRes}, the vapor phase is supersaturated  so that the density is higher 
than the corresponding bulk value at saturation.
Nevertheless, the bulk values are employed for the definition of the drop boundary because 
the location of the interface is rather insensitive to the vapor density.
A sphere is fitted to the liquid--vapor interface, considering distances to the wall larger than $2~\sigma_f$ 
whereas no weighting factors are introduced in the fitting procedure.
The region close to the wall is excluded because it shows perturbations due to strong ordering effects.
The tangent on this sphere at the intersection with the wall ($y = 0$) is used to determine the contact angle (cf. \ref{fig:SampledDensProfile2D}).
The mean contact angle is determined from the density profile averaged over the entire production period.
The uncertainty is estimated by the standard deviation of contact angles evaluated every 500 000 time steps 
during the production period.

The interaction of a fluid particle with the wall is the cumulative interaction of that fluid particle 
with all wall particles \cite{Hamaker1937}.
This cumulative potential $u^\mathrm{\Sigma}$ depends on the density of the wall and the distance of the particle to the wall $y$.
As the wall potential is not uniform but periodic it also depends on the lateral position above the wall, 
given by $x$ and $z$.
At a given lateral position $(x,z)$ of the fluid particle, there is a minimum of this cumulative potential $u^\mathrm{\Sigma}_{min}(x, z)$
with respect to the distance $y$ from the wall. 
The average minimum potential 
\begin{equation}
 - W = \frac{1}{L_x  L_z} \int \limits_{0}^{L_x}{\int \limits_{0}^{L_z}{dx~dz~u^\mathrm{\Sigma}_{min}(x, z)}},
 \label{eq:minPotSurf}
\end{equation}
where $L_x$ and $L_z$ denote the system size in lateral dimensions, is defined by the average over these minima.
For the LJTS potential used in the present study, $W$ depends linearly on $\zeta$ via Eq. \eqref{eq:unlikeInteraction}.
While different measures of the solid--fluid interaction are possible\cite{Grzelak2010a,Forte2014}, in the present study 
the magnitude of $W$ is employed as a measure, following Grzelak et al. \cite{Grzelak2010a}.
The calculation of the surface of minimal potential is numerically accomplished by setting up 
a cubic mesh 
with a spacing of $\Delta x = \Delta z = 0.031 ~\sigma_f$ and $\Delta y = 0.01~\sigma_f$.
The average minimum potential of the standard wall investigated in the present study ($\rho_{s} = 1.07 ~\sigma_s^{-3}$) is given by 
\begin{equation}
W = 3.08~\zeta k T_c, 
\label{eq:AvMinPot}
\end{equation}
where $k$ is the Boltzmann constant and $T_c = 1.078 ~\epsilon_f / k$ is the critical temperature of the LJTS fluid\cite{Vrabec2006}.
The average minimum potential is given by $W = 4.83~\zeta k T_c$ for $\rho_s = 2.10~\sigma_f^{-3}$,
and $W = 8.07~\zeta k T_c$ for $\rho_s = 4.02~\sigma_f^{-3}$.
In the range of the solid density investigated in the present study, the average minimum potential correlates linearly 
with the solid density $\rho_s$.
$W$ is well described by Eq. \eqref{eq:WOverRhoS} (cf. supporting information):
\begin{equation}
  W(\zeta, \rho_s)  = \left[ 1.7 \frac{\rho_s }{ \sigma_f^{-3}} + 1.3 \right] \zeta k T_c.
 \label{eq:WOverRhoS}
\end{equation}
On average, the minimum potential is located $0.96 ~\sigma_f$ above the topmost wall layer. 
The topography of the surface of minimal potential along with the local potential values is shown in the supporting information.
According to Grzelak et al. \cite{Grzelak2010a}, the molecular roughness of the atomistically--resolved wall 
does not influence the contact angle.
As chemical heterogeneities as a second source for contact angle hysteresis \cite{deGennes85} are absent, it can be assumed here
that no hysteresis occurs.
Thus, the results of this study represent the true thermodynamically stable contact angle.

\section{Simulation Results \label{secSimRes}}
\subsection{Size Effects}
To test the influence of the system size on the contact angle and the validity of the present results, 
simulations with different numbers of fluid particles at otherwise constant conditions are performed.
The simulations are carried out for $T = 0.8~\epsilon_f / k$ and a reduced solid--fluid interaction energy $\zeta$  of 0.35, 0.5 and 0.65,
with numbers of fluid particles $N = 750$, 1500, 45000 and 150000.

\begin{figure}[ht!]
  \includegraphics[width = 8.333cm]{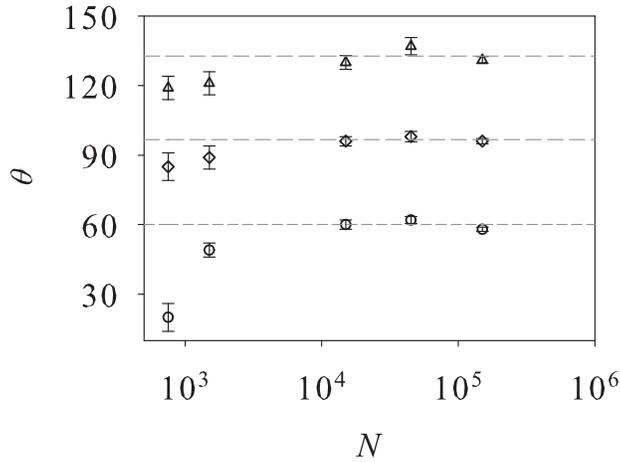}
 \caption{Impact of the number of fluid particles $N$ in the simulation box on the contact angle $\theta$ at $T = 0.8~\epsilon_f/k$ and 
	  reduced solid--fluid interaction energies $\zeta$ of 
	  0.35 ($\triangle$),
	  0.5 ($\Diamond$) and
	  0.65 ($\circ$).
	  The dotted lines correspond to the average of the results for the three largest particle numbers $N$.}
 \label{fig:SizeEff}
\end{figure}

The results are shown in \ref{fig:SizeEff}.
It can be seen that above about $N=10000$ fluid particles the observed contact angles do not depend significantly on the system size.
For the smaller system sizes, smaller contact angles are observed, consistently.
The deviation increases with increasing solid--fluid interaction energy.
There are several reasons for this deviation which, however, can not be identified seperately from the deviation of the contact angle \cite{Weijs2011}.
As can be seen in \ref{fig:SampledDensProfile2D}, there is a layering effect of the fluid density in vicinity to the wall.
For small droplets with $N = 750$ and 1500 particles, 
the layering affects the liquid density in the entire droplet and there are no bulk liquid properties.
\cite{Werth2013} found for planar liquid interfaces a significant decrease in the liquid--vapor interfacial tension 
due to the absence of bulk liquid properties, which is beyond the Tolman correction to the interfacial tension\cite{Tolman1949}.
In addition to the decrease of the liquid--vapor interfacial tension, the solid--liquid interfacial tension is assumed to decrease
by the lack of bulk liquid properties.
Another contribution affecting the contact angle is due to the growing influence of the three phase contact line 
with the line tension $\mathcal{L}$ and the curvature $\kappa$.
The influence of these effects can be assessed from\cite{Pethica1977} 
\begin{equation}
  \cos \theta = \frac{ \gamma_{sv} -  \gamma_{sl} }{\gamma_{lv} } - \frac{\mathcal{L} \kappa}{\gamma_{lv}},
 \label{eq:YoungExtLineTens}
\end{equation}
i.e. an adequately extended version of the Young equation
\begin{equation}
  \cos \theta = \frac{\gamma_{sv} - \gamma_{sl}}{\gamma_{lv}},
\label{eq:YoungsLaw}
\end{equation}
which both show that a decrease in $\gamma_{lv}$ would lead to a deviation such that 
the contact angle would be lower in the range of acute angles but higher for obtuse contact angles
(or vice versa, depending on the sign of the line tension).
The decrease in $\gamma_sl$ also contibutes to a persistent decrase in the contact angle.
The findings of the present study are in line with those of Santiso et al. \cite{Santiso2013}:
They observed a larger contact angle with an increasing droplet size and also a convergence to a constant angle.
In their case the contact angle converged at a system size of about $10^5$ fluid particles which is larger than in the present case.
This shift towards larger fluid particle numbers is attributed to the slower decay of their interaction potential as compared to the one used here.

\subsection{Density Profile \label{subsecDensProf}}
\begin{figure}[ht!]
\includegraphics[width = 8.333cm]{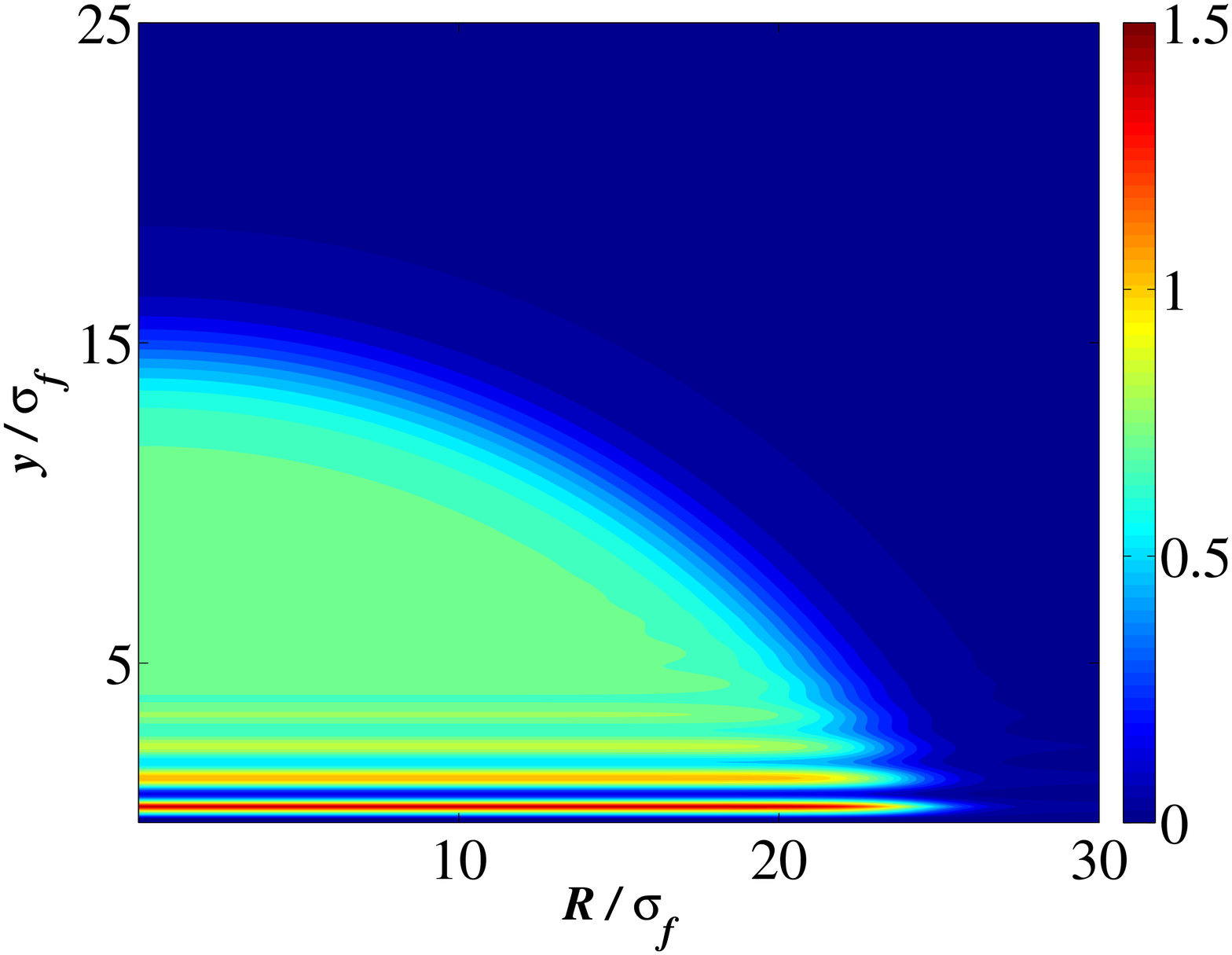}
\includegraphics[width = 8.333cm]{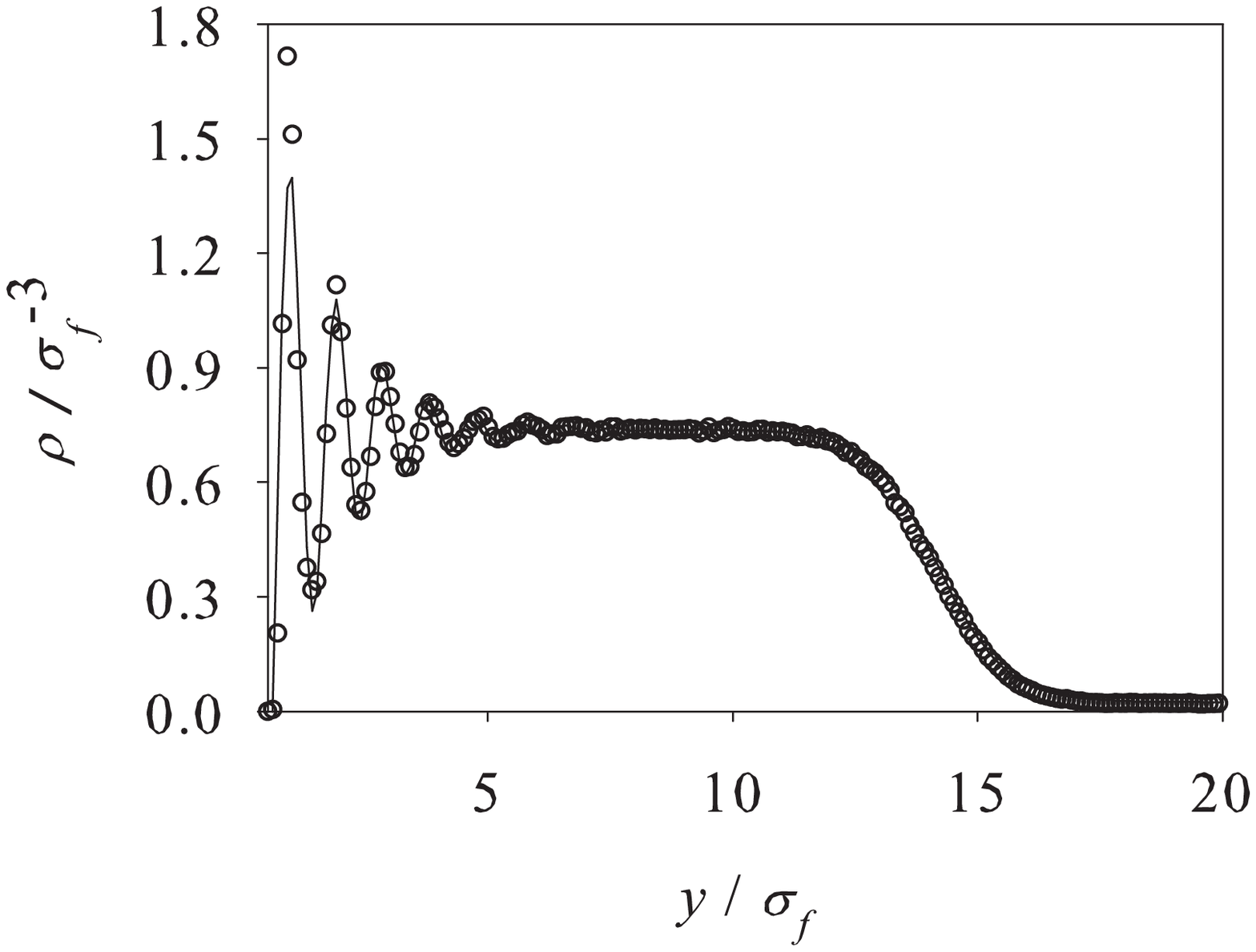}
\caption{Left: Correlation, cf. Eq.~\eqref{eq:densityCorrlationFnk}, of the density profile of a liquid drop adjusted to the data 
	 shown in \ref{fig:SampledDensProfile2D}, i.e. for a sessile drop at $\zeta = 0.65$ 
	 and $T = 0.8~\epsilon_f/ k$. 
	 Left: Two dimensional profile as obtained by the Eq. \eqref{eq:densityCorrlationFnk}.
	 The color indicates the density in units of $\sigma_f^{-3}$.
	 Right: Simulation results~($\circ$) and correlation ~(--), cf Eq.~\eqref{eq:densityCorrlationFnk},  for the density profile along
	 the axis of symmetry of the droplet ($R=0$) over distance from the wall.}
\label{fig:SampledDensProfile1D}
\end{figure}

For the case of $N = 15000$ fluid particles, the characteristics of the density profile of the fluid phase are studied 
at different values of the reduced solid--fluid interaction energy and the temperature. 
\ref{fig:SampledDensProfile2D} shows a typical density profile of the fluid phase.
The sessile drop on the planar wall has the shape of a spherical cap (circular in the two dimensional plot) and 
in the interfacial region, the density decreases radially from the center of the sphere to the vapor phase value.
The typical undulations in density due to the presence of the wall perturb the fluid only in a range of about 5 to $8~\sigma_f$. 
The essential features of the liquid and the vapor phase are correlated by the empirical ansatz 
\begin{equation}
\begin{split}
 \rho(\mathcal{R},y) & = f(\mathcal{R}) g(y), \\
 f(\mathcal{R}) & = \frac{1}{2} \left( \rho' + \rho'' \right) - \frac{1}{2}\left( \rho' - \rho'' \right) \tanh \left( \frac{2 (\mathcal{R}-\mathcal{R}_e)}{D} \right),  \\
 g(y) & = 1 + A \sin \left( \left( \frac{ y }{p} - s \right) 2 \pi \right) \exp \left( -c y  \right), 
\end{split}
\label{eq:densityCorrlationFnk}
\end{equation}
where $f(\mathcal{R})$ is the conventional function describing the density profile of a liquid drop surrounded 
by its vapor phase \cite{Rowlinson02}, with the liquid and vapor densities $\rho'$ and $\rho''$, respectively. 
The radius of the drop is $\mathcal{R}_e$ and the interfacial thickness is $D$.
Similarly, the sessile drop is considered as having a spherical shape, so that the density varies with radial distance $\mathcal{R}$ 
from the origin of the sphere.
The undulations of the fluid density in vicinity to the wall are modeled by a sinusoidal oscillation term with an amplitude $A$ and a period $p$.
The damping parameter $c$ characterizes the exponential decay of these undulations in terms of the distance from the wall $y$.
There are eight parameters $\rho'$, $\rho''$, $\mathcal{R}_e$, $D$, $A$, $p$, $s$ and $c$ that are determined from fitting 
the correlation \eqref{eq:densityCorrlationFnk} for each profile.
The numerical values of the parameters are given in the supporting information.
\ref{fig:SampledDensProfile1D} shows the correlation in cylindrical coordinates (left) and along the symmetry axis of the droplet (right) 
for the case of $\zeta = 0.65$ and $T = 0.8~\epsilon_f / k$.
In \ref{fig:SampledDensProfile1D}, $R$ denotes the distance from the symmetry axis of the drop, i.e. a cylindrical coordinate, 
as opposed to the spherical coordinate $\mathcal{R}$ from Eq.~\eqref{eq:densityCorrlationFnk}.

The correlation performs best for intermediate values of the reduced solid--fluid interaction energy for which the contact angle ranges 
from approximately $45^\circ$ to $135^\circ$.
Beyond that range, the correlation shows significant deviations from the densities observed in the simulation.
The curvature of the liquid--vapor interface yields vapor densities that differ from the saturation densities in the planar case 
due to the additional Laplace pressure \cite{Rowlinson02}.
The densities that are obtained from Eq.~\eqref{eq:densityCorrlationFnk} are compared to the saturation densities 
which are obtained by the conditions of phase equilibrium for curved interfaces.
A more detailed description can be found in the supporting information.
Fair agreement between the fluid densities obtained via Eq.~\eqref{eq:densityCorrlationFnk} and the saturation densities 
for the curved interface is found.
The average deviation of the liquid density is $2.2~\%$ and that of the vapor density is $6.9~\%$.
The interfacial thickness $D$ varies between $2.3~\sigma_f$ at $T = 0.7~\epsilon_f/k$ and about $8~\sigma_f$ at $T=1.0~\epsilon_f/k$ 
which agrees with the results of Vrabec et al. \cite{Vrabec2006}.
The period $p$ of the density undulations is found to be about $0.9~\sigma_f$ throughout, as it is characteristic for a packing structure.
The damping parameter $c$ of the density undulations increases from $0.5~\sigma_f^{-1}$  to $3.0~\sigma_f^{-1}$ 
at elevated temperatures and low values of the reduced solid--fluid interaction energy.
It corresponds to a decay length of about $2~\sigma_f$ at low temperatures and strong interaction to $0.3~\sigma_f$ 
at high temperatures and weak interaction, respectively.
The radius $\mathcal{R}_e$ of the drop from Eq.~\eqref{eq:densityCorrlationFnk} agrees well with the dividing surface 
that is determined by the threshold $\rho = \left( \rho' + \rho''\right)/2$.
As the interfacial thickness and the density undulations are independent from the system size\cite{Brovchenko12}, 
the density profile can be extrapolated to droplets of different size.

\subsection{Contact Angle}

\begin{figure}[ht!]
 \includegraphics[width = 8.333cm]{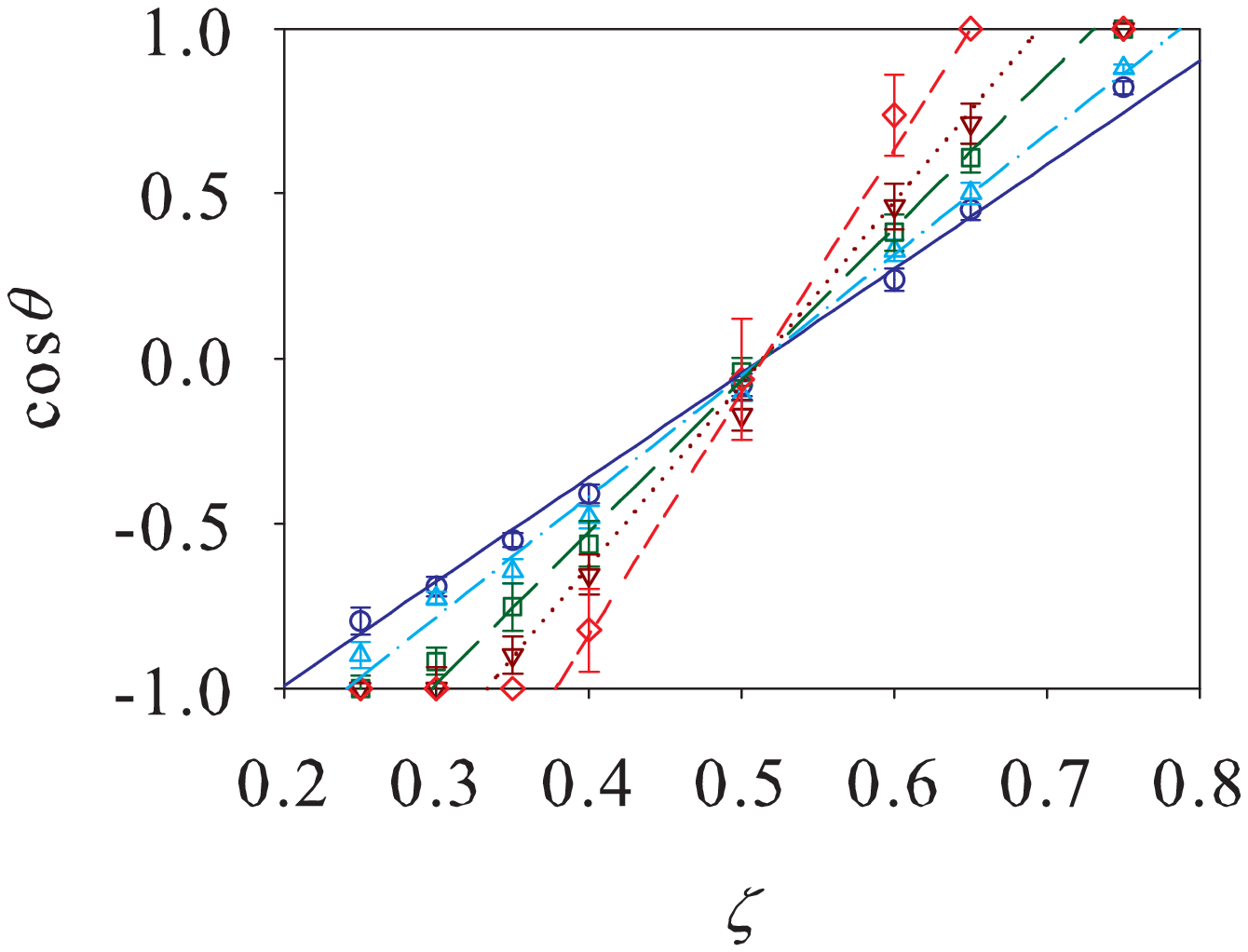}
 \caption{Simulation results (symbols) and correlation (lines), cf. Eq.~\eqref{eq:eqCorrelationCA}, for the contact angle 
 as a function of the reduced solid--fluid interaction energy at temperatures of $T = $ 
  $0.7$ ($\circ$, \hdashrule[0.5ex][x]{1.0cm}{0.8pt}{}),
  $0.8$ ($\triangle$, \hdashrule[0.5ex][x]{1.2cm}{0.8pt}{2.5mm 3pt 0.5mm 3pt}),
  $0.9$ ($\square$, \hdashrule[0.5ex][x]{1.3cm}{0.8pt}{5mm 3pt}),
  $0.95$ ($\triangledown$, \hdashrule[0.5ex][x]{1.2cm}{0.8pt}{2pt}) and
  $1.0~\epsilon_f/k$ ($\Diamond$, \hdashrule[0.5ex][x]{1.2cm}{0.8pt}{2.5mm 3pt 2.5mm 3pt 2.5mm 3pt}).
  The wall density is $\rho_s = 1.07~\sigma_f^{-3}$, i.e. $\sigma_s = \sigma_f$.}
  \label{fig:CAcorrelation1varxi}
\end{figure}

\begin{figure}[ht!]
  \includegraphics[width = 8.333cm]{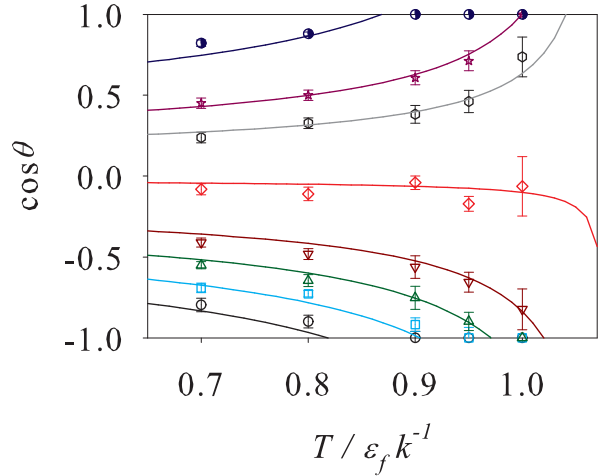}
 \caption{Simulation results (symbols) and correlation (lines), cf. Eq.~\eqref{eq:eqCorrelationCA}, for the contact angle 
 as a function of temperature at values of the solid--fluid interaction energy $\zeta$ of
 $0.25$ ($\circ$),
 $0.3$ ($\square$),
 $0.35$ ($\triangle$), 
 $0.4$ ($\triangledown$),
 $0.5$ ($\Diamond$),
 $0.6$ ($\hexagon$),
 $0.65$ ($\star$) and 
 $0.75$ ($\RIGHTcircle$).
 The wall density is $\rho_s = 1.07~\sigma_f^{-3}$, i.e. $\sigma_s = \sigma_f$.}.
 \label{fig:CAcorrelation1varT}
\end{figure}

The reduced solid--fluid interaction energies are varied at temperatures between 0.7 and 1.0 $\epsilon_f/ k$. 
This covers most of the range of the vapor--liquid coexistence of the LJTS fluid 
between the triple point\cite{Meel2008} at the temperature  $0.65 ~\epsilon_f/ k$ and the critical point\cite{Vrabec2006} 
at the temperature $1.078 ~\epsilon_f / k$.
The simulation results are shown in \ref{fig:CAcorrelation1varxi} and \ref{fig:CAcorrelation1varT}, 
and the corresponding numerical data are listed in the supporting information.
For values of the reduced solid--fluid interaction  between 0.25 and 0.75, the contact angle varies from  
total wetting to total dewetting (i.e. $0^\circ \leq \theta \leq 180^\circ$).
The correlation 
\begin{equation}
\cos\theta(\tau,\zeta) = \alpha(1+ \tau^{\delta})(\zeta - \zeta_0) \text{,}
\label{eq:eqCorrelationCA}
\end{equation}
where $\tau = (1-T/T_c)$, was adjusted to the simulation results and yields good agreement for 
$\alpha = 1.03$, $\delta = -0.69$,  and $\zeta_0 = 0.514$, cf. \ref{fig:CAcorrelation1varxi}.
While it is not fully resolved wheter the nature of the drying transition is first or second order in the case of short range potentials
\cite{Nijmeijer1990, Henderson1990}, the linear correlation was chosen as it shows only minor differences to the simulation data.
The parameter $\delta$ that characterizes the temperature dependence of the contact angle is assumed to be independent from 
the solid--fluid potential and to solely depend on the temperature.
Thus, it is fixed to $\delta  = -0.69$ throughout.
Due to the linear relation between the average minimum potential and the reduced solid--fluid interaction energy $\zeta$ (Eq. \eqref{eq:AvMinPot}),
this transforms to
\begin{equation}
\cos\theta(\tau,W) = \frac{\bar{\alpha}}{k T_c}(1+ \tau^{\delta})(W - W_0),  
\label{eq:eqCorrelationCAReduced}
\end{equation}
with $\bar{\alpha} = 0.335$.
The contact angle $\theta = 90^\circ$ occurs at $\zeta_0$ 
and thus at an average minimum potential given by $W_0 = W (\zeta_0) = 1.58~k T_c$.
The value of $\zeta_0$, and hence that of $W_0$, are both found to be independent of temperature, confirming previous work \cite{Horsch2010}.
High values of $\zeta$ correspond to a strong attraction between the fluid and the wall.
As expected, an increasing solid--fluid attraction leads to a decreasing contact angle (see \ref{fig:CAcorrelation1varxi}).

Eqs. \eqref{eq:eqCorrelationCA} and \eqref{eq:eqCorrelationCAReduced} confirm the symmetry relation
\begin{equation}
 \cos \theta(\tau, \zeta_0 + \Delta \zeta  ) = - \cos\theta(\tau, \zeta_0 - \Delta \zeta )
 \label{eq:CASymmetry}
\end{equation}
previously found by Horsch et al. \cite{Horsch2010}, Sikkenk et al. \cite{Sikkenk1988} as well as Monson\cite{Monson2008} both by MD simulations 
and by DFT calculations.

The Young equation \cite{Young1805}, cf. Eq.~\eqref{eq:YoungsLaw}, 
relates the contact angle $\theta$ to the interfacial tensions $\gamma_{sv}$, $\gamma_{sl}$ and $\gamma_{lv}$ of the 
solid--vapor, solid--liquid and liquid--vapor interfaces, respectively.
The liquid--vapor interfacial tension $\gamma_{lv}$ is not affected by the solid--fluid interaction.
At a given temperature, it follows from Eqs. \eqref{eq:eqCorrelationCA} and \eqref{eq:YoungsLaw} that 
the interfacial tension difference $\gamma_{sv} - \gamma_{sl}$ varies linearly with 
the reduced solid--fluid interaction energy $\zeta$.
At $\theta = 90^\circ$ the interfacial tensions $\gamma_{sv}$ and $\gamma_{sl}$ are equal.
The present simulation  results indicate that the conditions for which $\gamma_{sv} = \gamma_{sl}$ 
is fulfilled do not depend on the temperature.

The influence of the temperature on the contact angle is shown in \ref{fig:CAcorrelation1varT}:
The extent of wetting or dewetting increases at elevated temperatures.
Applying the Young equation indicates that the ratio of $|\gamma_{sv} - \gamma_{sl}| / \gamma_{lv}$ increases at higher temperatures.
A transition occurs when $| \gamma_{sv} - \gamma_{sl}|  = \gamma_{lv}$.
For the case of a contact angle of $180^\circ$, an intrusion of a solid--vapor interface 
below the droplet is observed, cf. \ref{fig:SampledDensProfile2D} (right).

\subsection{Wall Density}
\begin{figure}[ht!]
\includegraphics[width = 8.333cm]{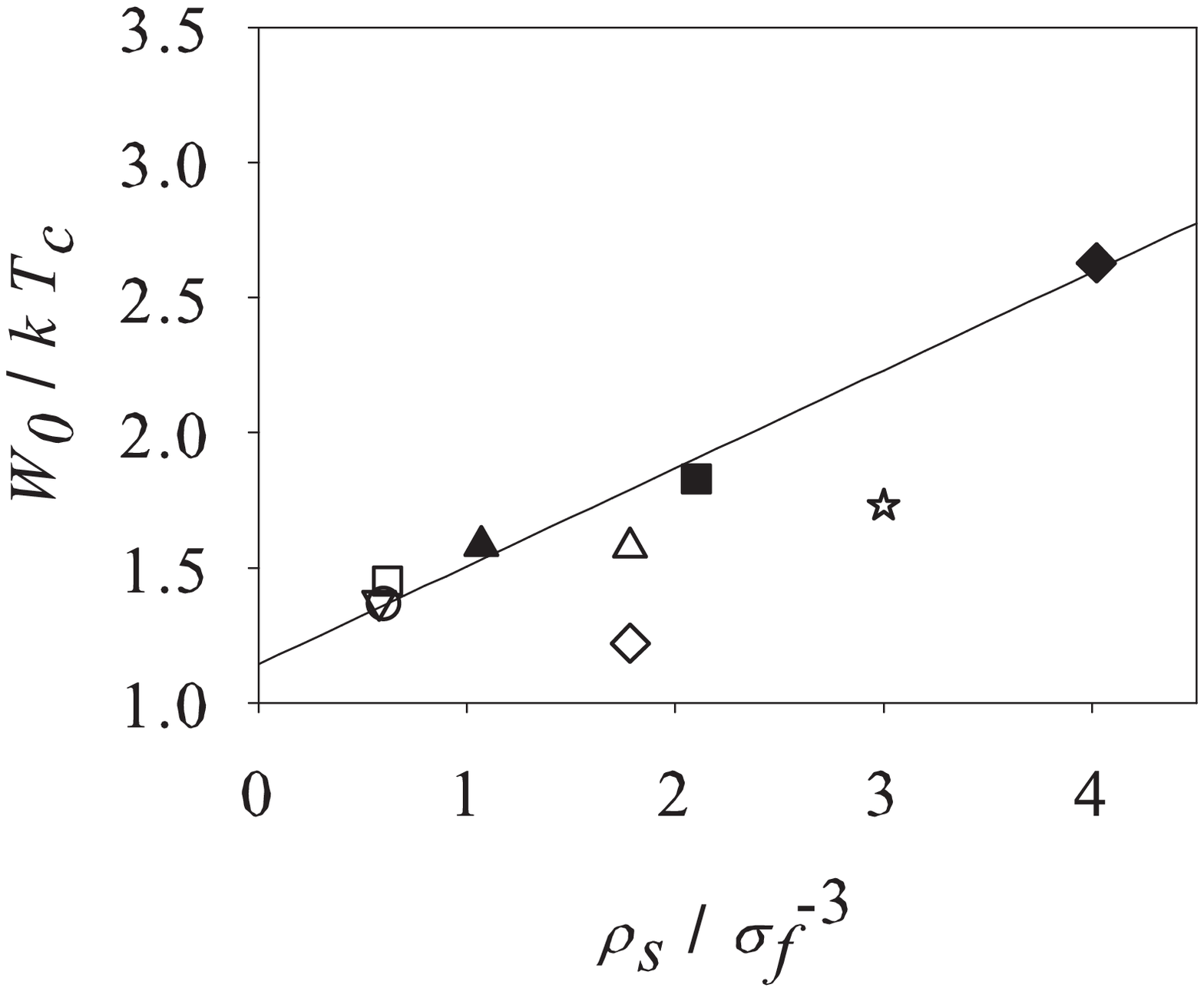}
\includegraphics[width = 8.333cm]{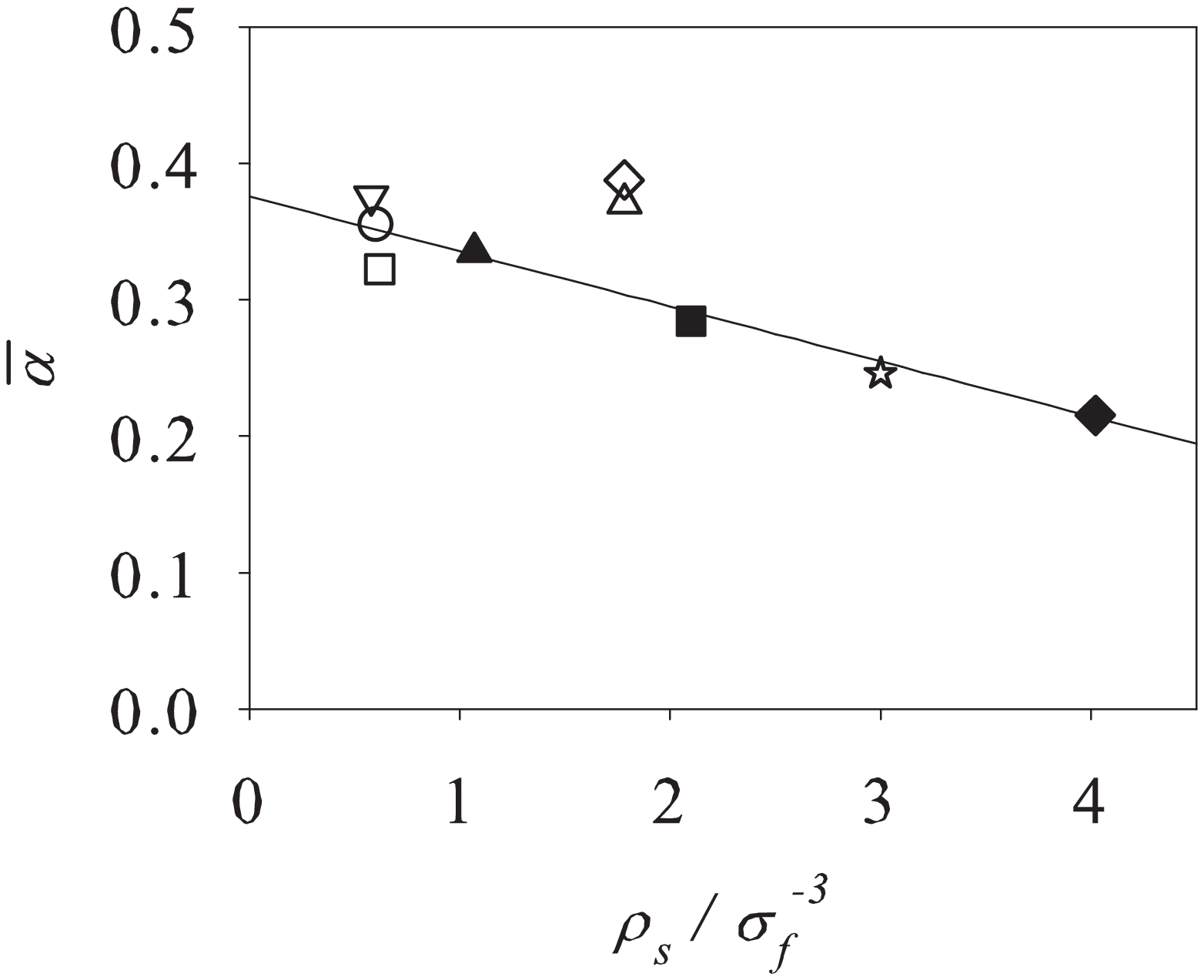}
\caption{ Left: Value of $W_0$, corresponding to the average minimum potential that yields $\theta = 90^\circ$, for wall models of different density.
	  Right: Gradient $\bar{\alpha}$ of the contact angle cosine over the solid density.
	  This work: $\rho_s = 1.07 \sigma_f^{-3}$ ($\blacktriangle$), 
		     $\rho_s = 2.10 \sigma_f^{-3}$ ($\blacksquare$), 
		     $\rho_s = 4.02 \sigma_f^{-3}$ ($\Diamondblack$). 
	 Ingebrigtsen and Toxvaerd \cite{Ingebrigtsen2007} ($\Circle$);
	 Shaharaz et al. \cite{Shahraz2012} (\ding{73});
	 Grzelak et al. \cite{Grzelak2010a}, bcc~(100) lattice ($\triangledown$);
	 Tang and Harris \cite{Tang1995} ($\square$);
	 Nijmeijer et al. \cite{Nijmeijer1990} ($\triangle$);
	 Nijmeijer et al. \cite{Nijmeijer1992} combined model ($\Diamond$);
	 The lines represent the fit obtained using from the present simulation data using Eqs.~\eqref{eq:W0Linear} and \eqref{eq:alphaBarOverRhoWall}.
	 The parameters are $\eta_1 = 0.36$, $\eta_2 = 1.1$, $\eta_3 = -0.04$, and $\eta_4 = 0.38$.}
\label{fig:W0OverRhoS}
\end{figure}

In order to study the influence of the solid density on the contact angle, simulations are carried out not only for a wall of 
the density of $\rho_s = 1.07~\sigma_f^{-3}$ (results discussed above) but also for walls of two other
densities: $\rho_s = 2.10~\sigma_f^{-3}$ and  $\rho_s = 4.02~\sigma_f^{-3}$.
The simulation results for the contact angles on surfaces with increased solid densities are correlated 
by Eq. \eqref{eq:eqCorrelationCAReduced} using the same value for $\delta = -0.69$ as given above, 
but newly adjusted values for the average minimum potential at $\theta = 90^\circ$, which is given by $W_0$, and the gradient $\bar{\alpha}$.
The numerical results are shown in the supporting information.
In particular, both $W_0$ and $\bar{\alpha}$ are found to depend linearly on the solid density $\rho_s$.
Correlations for $W_0(\rho_s)$ and $\bar{\alpha}(W_0)$ are obtained by fitting expressions
\begin{equation}
 W_0(\rho_s)/ k T_c = \eta_1 \frac{\rho_s }{ \sigma_f^{-3}} + \eta_2 ,
 \label{eq:W0Linear}
\end{equation}

\begin{equation}
\bar{\alpha}(\rho_s) = \eta_3 \frac{\rho_s }{ \sigma_f^{-3}} + \eta_4,
 \label{eq:alphaBarOverRhoWall}
\end{equation}
to the present simulation results.
Good results are obtained for $\eta_1 = 0.36$, $\eta_2 = 1.1$, $\eta_3 = -0.04$, and $\eta_4 = 0.38$ (cf. \ref{fig:W0OverRhoS}).

\section{Discussion\label{secDiscussion}}

Contact angles in LJ systems have been studied by different authors before. 
\ref{tab:ComparativeModelDetails} gives an overview in which also the results of the present study are summarized. 
There are two additional studies: One by Bucior et al.\ \cite{Bucior2009} who have investigated systems with 
only a single layer of wall interaction sites, arranged in a closest hexagonal packing.
In the study of Horsch et al. \cite{Horsch2010}, the wall model was meant to represent graphite.
Both wall models are characterized by a high lateral density.
In the case of the graphite model, the interlayer distance is large (about $0.9~\sigma_f$).
The arrangement of the solid sites in both studies was forced, and densities vary significantly from the equilibrium configuration for 
a solid interacting via a LJ potential. 
The potential characteristics will therefore be different from those of the other studies.
Accordingly, their results are not quantitatively comparable to the other investigations (e.g. see the data of Horsch et al.\cite{Horsch2010} 
in Figure~S.3 in the supporting information).
Therefore, the studies of Bucior et al. \cite{Bucior2009} and Horsch et al. \cite{Horsch2010} are not further discussed here.
Furthermore, there are studies that basically mimic one of the models discussed here for the purpose of comparison\cite{Leroy2010,Rane2011}.
They are not considered in the present discussion, either.

The solid--fluid potential of the literature models differ both in the potential type and the cutoff radius.
Ingebrigtsen and Toxvaerd \cite{Ingebrigtsen2007} have used a continuous LJ 9--3 potential representing 
the cumulative interaction of a fluid particle with the wall.
Shahraz et al.\cite{Shahraz2012} have also used a contiuous LJ 9--3 potential that differs from the model of Ingebrigtsen and Toxvaerd \cite{Ingebrigtsen2007} 
in the ineraction strength. 
Furthermore, Shahraz et al. \cite{Shahraz2012} consider a simulation setup where they investigate the contact angle of an infinitely long cylindrical
LJ droplet.
All other studies mentioned here consider droplets assuming the shape of a spherical cap. 
Nijmeijer et al. \cite{Nijmeijer1992} have used a combination of a particulate and continuous LJ~9--3 solid--fluid potential, 
whereas all other authors \cite{Nijmeijer1990, Grzelak2010a,Tang1995} have used particulate models.
The solid density was similar for the studies of Ingebrigtsen et al. \cite{Ingebrigtsen2007}, Tang and Harris \cite{Tang1995}, 
and Grzelak et al. \cite{Grzelak2010a} ($\rho_s \approx 0.6~\sigma_f^{-3}$).
Grzelak et al. \cite{Grzelak2010a} have studied the contact angle on several wall models at a constant solid density 
$\rho_s = 0.58~\sigma_f^{-3}$, but for various lattice structures and surface orientations.
They found a strong correlation between the average minimum potential and the contact angle, i.e. 
the contact angle was well characterized by the average minimum potential.
For that reason, only one of the wall models of that literature source is discussed in the present study, 
namely the body centered cubic (bcc) wall with the (100) surface exposed to the fluid.
In the following, it is referred to as the ``bcc (100) lattice''. 
Furthermore, there are several closely related MD studies on wetting in a LJ system by Sikkenk et al. \cite{Sikkenk1987, Sikkenk1988}
as well as Nijmeijer et al \cite{Nijmeijer1989, Nijmeijer1990, Nijmeijer1992}.
All these studies use very similar molecular models and scenarios.
The present discussion exemplarily refers to two of those studies, both by Nijmeijer et al.\cite{Nijmeijer1990, Nijmeijer1992}.
In those simulations, the solid density was $\rho_s = 1.78~\sigma_f^{-3}$.
In the first study\cite{Nijmeijer1990}, the solid--fluid potential was particulate.
The other simulation study discussed here\cite{Nijmeijer1992} used a particulate solid--fluid potential and an additional cutoff correction 
in form of a LJ~9--3 potential.
This is referred to as the Nijmeijer et al.\cite{Nijmeijer1992} combined  model, in the following.
It is similar to the one used by Ingebrigtsen and Toxvaerd \cite{Ingebrigtsen2007} and it was meant to account for the long range contribution of
the LJ potential.
However, while the Nijmeijer et al.\cite{Nijmeijer1992} combined  model does consider a long--range correction 
contribution to the forces acting on fluid particles at distances $y > r_c$ from the wall, the long--range forces are completely neglected 
close to the wall ($y \leq r_c$).
Thereby, the Nijmeijer et al. \cite{Nijmeijer1992} combined model, which is considered here nonetheless, fails to consistently address the issue 
of scale separation, since both short--range and long--range forces are actually strongest close to the wall.
The way this combined potential was implemented therefore seems to be inconsistent to the present authors.

The studies of the different authors are carried out at constant but different temperatures. 
For some of the studies\cite{Ingebrigtsen2007,Grzelak2010a, Shahraz2012} mentioned above, the average minimum potential $W$
could be directly obtained from the the literature source.
For the studies of Nijmeijer et al. \cite{Nijmeijer1990, Nijmeijer1992} as well as Tang and Harris \cite{Tang1995}, 
the walls were reconstructed and the average minimum potential was calculated using Eq. \eqref{eq:minPotSurf}.
It may be noted that for the earlier study of Nijmeijer et al. \cite{Nijmeijer1990}, the corrected value of the solid--fluid 
cutoff radius of $2.21~\sigma_f$ was used, as it was reported in the erratum in the subsequent paper by Nijmeijer et al. \cite{Nijmeijer1992}.
The simulation data for the contact angles from the literature were fitted using Eq. \eqref{eq:eqCorrelationCAReduced} with $\delta = -0.69$.
The results of the fit for the two correlation parameters $W_0$ and $\bar{\alpha}$, that were adjusted to the literature data,
are included in \ref{tab:ComparativeModelDetails}.
They are very well predicted by the correlation obtained from the simulation data of the present work,
cf. Eqs. \eqref{eq:W0Linear} and \eqref{eq:alphaBarOverRhoWall}.
The results of the Nijmeijer et al.\cite{Nijmeijer1992} combined model, however, deviate considerably.

\begin{figure}[ht!]
 \includegraphics[width = 8.333cm]{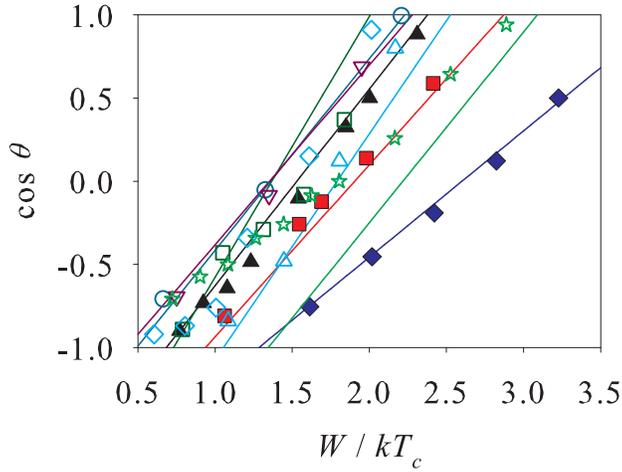}
 \caption{Contact angle cosine over the absolute magnitude $W$ of the average minimum potential.  
  The lines represent Eq.~\eqref{eq:CAPredicted} with the parameters $\eta_1 = 0.36$, $\eta_2 = 1.1$, $\eta_3 = -0.04$, and $\eta_4 = 0.38$.
  The symbols are the simulation results at different temperatures $T$. 
  This work at a temperature of $0.8 \epsilon_f / k$:
  $\rho_s = 1.07 \sigma_f^{-3}$ ($\blacktriangle$), 
  $\rho_s = 2.10 \sigma_f^{-3}$ ($\blacksquare$), 
  $\rho_s = 4.02 \sigma_f^{-3}$ ($\Diamondblack$).
  Ingebrigtsen and Toxvaerd \cite{Ingebrigtsen2007}, $0.75~\epsilon_f / k$, ($\Circle$);
  Shaharaz et al. \cite{Shahraz2012}, $0.7~\epsilon_f / k$, (\ding{73});
  Grzelak et al. \cite{Grzelak2010a}, bcc~(100) lattice, $0.70~\epsilon_f / k$, ($\triangledown$);
  Tang and Harris \cite{Tang1995}, $0.90~\epsilon_f / k$,($\square$);
  Nijmeijer et al. \cite{Nijmeijer1990}, $0.90~\epsilon_f / k$, ($\triangle$);
  Nijmeijer et al. \cite{Nijmeijer1992} combined model ($\Diamond$).
  The simulation results and the corresponding lines representing Eq.~\eqref{eq:CAPredicted} are linked by the colors.}
 \label{fig:CAPredicted}
\end{figure}

The results from the correlations obtained in the present study can furthermore directly be compared to the simulation data 
for the contact angle from the different sources.
The contact angles are predicted by Eq. \eqref{eq:eqCorrelationCAReduced}  using solely the information on the temperature 
and the solid density given in the literature sources.
The solid density was used to determine $\bar{\alpha}(\rho_s)$ and $W_0(\rho_s)$ via Eqs. \eqref{eq:W0Linear} and \eqref{eq:alphaBarOverRhoWall}. 
The final correlation has the form
\begin{equation}
  \cos\theta(\tau,W,\rho_s) = \left(\eta_1 \frac{\rho_s}{\sigma_f^{-3}} + \eta_2\right) \left(1+ \tau^{\delta}\right) 
			    \left[ W - \left( \eta_3 \frac{\rho_s }{ \sigma_f^{-3}} + \eta_4\right) \right] \text{,}
 \label{eq:CAPredicted}
\end{equation}
with the parameters $\eta_i$, $i=1...4$ and $\delta = -0.69$ as introduced above.
As can be seen from \ref{fig:CAPredicted}, a good agreement is obtained for most of the simulation data from the literature sources.
The results of the Nijmeijer et al. \cite{Nijmeijer1992} combined model differ considerably from the prediction, which is 
attributed to the special type of the solid--fluid potential that was mentioned before.
Also, the contact angle data of the cylindrical droplet from the study of Shahraz et al. \cite{Shahraz2012} deviate.
This might be attributed to the different topology of their simulation setup.
The general agreement between Eq. \eqref{eq:CAPredicted} and the simulation data is also obtained for 
literature data that are not shown in \ref{fig:CAPredicted}, for clarity.

\section{Conclusions \label{secConclusions}}
Sessile drops on a solid wall were studied in a LJTS system.
The temperature, the wall density, and  the strength of the dispersive fluid--solid interaction were systematically varied.
Simulation results for the contact angle as a function of these parameters were obtained.
The present simulation data considerably extend the previously available information on systems of the studied type.
A correlation which describes the dependence of the contact angle on the parameters mentioned above was developed using 
the data from the present study.
This novel and general correlation agrees well with simulation data obtained by other authors in previous studies on the contact angle in LJ systems, 
even though details of the models differ.

\section{Acknowledgement}
The authors gratefully acknowledge financial support by the DFG within CRC 926
``Microscale Morphology of Component Surfaces''.
Computational support is acknowledged by 
the Leibniz Supercomputing Center (LRZ) under the large-scale grant pr83ri 
and the Regional University Computing Center Kaiserslautern (RHRK) under the grant TUKL--MSWS. 
The authors thank Tobias Alter for carrying out some of the simulations, 
as well as Cemal Engin and Jadran Vrabec for fruitful discussions.
The present work was conducted under the auspices of the Boltzmann--Zuse Society for 
Computational Molecular Engineering (BZS).

\section{Tables and Figures}


\begin{table}[ht!]
\caption{LJ model systems used for studies on wetting. The solid--fluid potential of 
Ingebrigtsen  and Toxvaerd \cite{Ingebrigtsen2007} and of Shaharaz et al. \cite{Shahraz2012} is a continuous LJ~9--3 model, 
all other potentials are particulate LJ~12--6 models.
The values of $W/\zeta$ were obtained from the literature sources (see text). 
The parameters $W_0$ and $\bar{\alpha}$ of a correlation of the literature data based on Eq. \eqref{eq:eqCorrelationCAReduced} are given.}
\begin{tabular}{ | c | c | c |c | c | c | c |}
\hline 
source 					& $\rho_s$/  		& $r_{c,sf}$/  		& $T$ /			&$W $ / 	& $W_0$ /	& $\bar{\alpha}$\\
~					& $\sigma_f^{-3}$ 	& $\sigma_f$		& $\epsilon_f / k$ 	&$\zeta k T_c$	& $k T_c$	& 		\\	
\hline
Ingebrigtsen    			& 0.60 			& $\infty$ 		& $0.75$		&$1.32$		& $1.37$	& $0.356$	\\
and Toxvaerd \cite{Ingebrigtsen2007}	& ~			& ~			& ~			&~		& ~		& ~		\\
\hline
Shaharaz et al. \cite{Shahraz2012}	& 3.0			& $\infty$		& $0.7$			&$3.61$		&$1.73$		&$0.246$		\\
\hline
Grzelak et al. \cite{Grzelak2010a} 	& 0.58			& $5.00$		& $0.70$		&$3.00$		& $1.38$	& $0.376$ 	\\
bcc(100) lattice			&~ 			&~ 			&~ 			& ~		&~ 		&~ 		\\	
\hline
Tang and Harris \cite{Tang1995}		& 0.62			& $ 2.75$		& $0.90$		&$2.62$		& $1.45$	& $0.323$	\\
\hline
Nijmeijer et al. \cite{Nijmeijer1990}	&1.78			& $2.21$		&$0.90$			&$3.61$		& $1.58$	& $0.371$ 	\\
\hline
Nijmeijer et al. \cite{Nijmeijer1992}	&1.78			& $2.35$		&$0.90$			&$4.03$		& $1.22$	& $0.338$ 	\\
combined model				& ~			& ~			& ~			& ~		& ~		& ~		\\
\hline					
~					& 1.07			& 2.50			& $0.80$		&$3.08$		& $1.58$ 	& $0.335$	\\
\cline{2-7}
this work				& 2.10			& 2.50			& $0.80$		&$4.83$		& $1.82$ 	& $0.288$	\\
\cline{2-7}
~					& 4.02			& 2.5			& $0.80$		&$8.07$		& $2.67$ 	& $0.215$	\\
\hline
\end{tabular}
\label{tab:ComparativeModelDetails}
\end{table}

\providecommand*\mcitethebibliography{\thebibliography}
\csname @ifundefined\endcsname{endmcitethebibliography}
  {\let\endmcitethebibliography\endthebibliography}{}

\newpage


\setcounter{equation}{0}
\renewcommand{\theequation}{S-\arabic{equation}}

\setcounter{table}{0}
\renewcommand*\thetable{S.\arabic{table}}

\setcounter{figure}{0}
\renewcommand*\thefigure{S.\arabic{figure}}

The saturation densities of the curved interfaces are determined by the conditions of phase equilibrium 
that are stated in Eqs.~\eqref{eq:phaseEquiCondT} to~\eqref{eq:phaseEquiCondMu}:
\begin{align}
\label{eq:phaseEquiCondT}
 T'   		&= T'' \\
 \label{eq:phaseEquiCondP}
 p'   		&= p'' + \Delta p\\
 \label{eq:phaseEquiCondMu}
 \mu'(T,p,N) 	&= \mu''(T,p,N),
\end{align}
where a single prime denotes the liquid phase and two primes denote the vapor phase.
The chemical potential is denoted by $\mu$, and the number of fluid particles of the single fluid component by N.

For a liquid drop, the pressure difference $\Delta p$ in Eq. \eqref{eq:phaseEquiCondP} is positive 
and is obtained from the Laplace equation\cite{Rowlinson02}, Eq.~\eqref{eq:Laplace}:
\begin{equation}
  \Delta p = \frac{2\gamma_{lv}}{\mathcal{R}_e}.
 \label{eq:Laplace}
\end{equation}
The interfacial tension data $\gamma_{lv}$ that are introduced in Eq.~\eqref{eq:Laplace} are taken from the results of Vrabec et al.\cite{Vrabec2006} 
for the LJTS fluid with a planar interface. 
The capillarity approximation is applied, i.e. the interfacial tension is assumed to be independent from the curvature of the interface. 
The radius $\mathcal{R}_e$ is obtained from Eq.(8).

At the given temperature and pressure difference, the phase equilibrium is obatained by equating the chemical potentials of the liquid 
and the vapor phase (see Eq.~\eqref{eq:phaseEquiCondMu}).
The chemical potential of the liquid phase is described by Eq.~\eqref{eq:ChemPotLiq}:
\begin{equation}
  \mu'(T,p) = \mu_0 + \int_{p_s}^{p'}v(T,p) dp
 \label{eq:ChemPotLiq}
\end{equation}
where $\mu_0$ is the chemical potential at the liquid--vapor coexistence of the planar interface which serves as a refernce point.
$p_s$ is the vapor pressure of the fluid with a planar interface and $p'$ is the pressure of the liquid drop according to Eq. \eqref{eq:phaseEquiCondP}.
The molar volume $v$ is obtained by the fifth order virial isotherm of Horsch et al.\cite{Horsch2010a} descirbing the 
saturation properties of the LJTS fluid at liquid--vapor phase coexistence.

The chemical potential of the vapor phase is described by Eq.~\eqref{eq:ChemPotVap}:
\begin{equation}
  \mu''(T,p) = \mu_0 + \int_{p_s}^{p''}v(T,p) dp
 \label{eq:ChemPotVap}
\end{equation}
where $p''$ is the pressure of the vapor phase according to Eq. \eqref{eq:phaseEquiCondP}.

Introducing Eqs.~\eqref{eq:Laplace}, \eqref{eq:ChemPotLiq}, and \eqref{eq:ChemPotVap} into Eqs.~\eqref{eq:phaseEquiCondT} 
to~\eqref{eq:phaseEquiCondMu}  yields the state point at liquid--vapor coexistence and hence the molar volumes $v'$ and $v''$.

\newpage

\begin{table}[hb]
\caption{Contact angle of the LJTS fluid on a solid wall from the present MD simulations with a wall density of $\rho_s = 1.07~\sigma_f^{-3}$.
	 The correlation using Eq.~(10) yields $W_0 =1.58~k T_c$ and $\bar{\alpha} = 0.335$.} 
\begin{tabular}{ | c ||r@{$\pm$}  l | r@{$\pm$} l | r@{$\pm$} l | r@{$\pm$} l | r@{$\pm$} l |}
\hline 
$\zeta$ 	&\multicolumn{10}{c|}{$kT/\epsilon_f$}\\
\hline
   & \multicolumn{2}{|c|}{$0.7$}    	& \multicolumn{2}{|c|}{$0.8$}   	& \multicolumn{2}{|c|}{$0.9$}  	& \multicolumn{2}{|c|}{$0.95$} 	& \multicolumn{2}{|c|}{$1.0$}\\
\hline \hline
0.25	& $143^\circ $ & $ 4^\circ$	& $154^\circ $ & $ 5^\circ$	& $180^\circ $ & $ 11 ^\circ$	& $180^\circ $ & $ 0.0^\circ$	& $180^\circ $ & $ 0^\circ$	\\
\hline
0.30	& $134^\circ $ & $ 2^\circ$	& $137^\circ $ & $ 2^\circ$	& $157^\circ $ & $ 7^\circ$	& $180^\circ $ & $ 14^\circ$	& $180^\circ $ & $ 0^\circ$	\\
\hline 
0.35	& $123^\circ $ & $ 2^\circ$	& $130^\circ $ & $ 3^\circ$	& $139^\circ $ & $ 6^\circ$	& $154^\circ $ & $ 9^\circ$	& $180^\circ $ & $ 0^\circ$	\\
\hline
0.40	& $114^\circ $ & $ 2^\circ$	& $119^\circ $ & $ 2^\circ$	& $124^\circ $ & $ 5^\circ$	& $131^\circ $ & $ 5^\circ$	& $145^\circ $ & $ 17^\circ$	\\
\hline 
0.50	& $95^\circ $ & $ 2^\circ$	& $96^\circ $ & $ 2^\circ$	& $92^\circ $ & $ 2^\circ$	& $100^\circ $ & $ 2^\circ$	& $94^\circ $ & $ 11^\circ$	\\
\hline 
0.60	& $76^\circ $ & $ 2^\circ$	& $71^\circ $ & $ 2^\circ$	& $68^\circ $ & $ 3^\circ$	& $63^\circ $ & $ 5^\circ$	& $43^\circ $ & $ 12^\circ$	\\
\hline 
0.65	& $63^\circ $ & $ 2^\circ$	& $60^\circ $ & $ 2^\circ$	& $53^\circ $ & $ 3^\circ$	& $45^\circ $ & $ 6^\circ$	& $0^\circ $ & $ 0^\circ$	\\
\hline 
0.75	& $35^\circ $ & $ 2^\circ$	& $28^\circ $ & $ 2^\circ$	& $0^\circ $ & $ 0^\circ$	& $0^\circ $ & $ 0^\circ$	& $0^\circ $ & $ 0^\circ$	\\
\hline
\end{tabular}
\label{tab:CAResults}
\end{table}
\newpage

\begin{table}[hp]
  \centering
  \caption{Simulation results for contact angles at a temperature of $T = 0.8~\epsilon_f / k$ and different wall 
	   densities $\rho_s$. The parameters $W_0$ and $\bar{\alpha}$ are obtained by a correlation based 
	   on Eq.~(10).} 
    \begin{tabular}{|c|c|r@{$\pm$} l| c | c | }
    \hline
    $\rho_s /$ 			& $\zeta$ 	& \multicolumn{2}{|c|}{$\theta$} 	& $W_0 /$	&$\bar{\alpha} / $\\
    $\sigma_f^{-3}$		& 		& \multicolumn{2}{|c|}{~}		& $ kT_c$ 	& \\
    \hline \hline
     ~				& 0.22  	& $144^\circ $	& $7^\circ$  		& ~ 		& ~\\
     \cline{2-4}
    ~				& 0.32  	& $105^\circ $ & $ 4^\circ$ 		& ~		& ~\\
    \cline{2-4}
    2.1				& 0.35  	& $97^\circ $ & $ 2^\circ$		& $1.82$ 	&$0.288$\\
    \cline{2-4}
    ~				& 0.41 		& $82^\circ $ 	& $ 3^\circ$ 		&  ~    	& ~\\
    \cline{2-4}
    ~				& 0.50   	& $54^\circ $ & $ 5^\circ$ 		& ~		& ~\\
    \hline
    ~				& 0.20 		& $139^\circ $ & $ 3^\circ$		& ~ 		& ~\\
    \cline{2-4}
    ~ 				& 0.25 		& $117^\circ $ & $ 1^\circ$		& ~		& ~\\
    \cline{2-4}
    4.02			& 0.30		& $101^\circ $ & $ 1^\circ$		& $2.63$	&$0.215$\\ 
    \cline{2-4}
    ~				& 0.35		& $83^\circ $ & $ 1^\circ$		& ~		& ~\\
    \cline{2-4}
    ~				& 0.40 		& $60^\circ $ & $ 2^\circ$		& ~		& ~\\
    \hline
    \end{tabular}%
  \label{tab:CAatDenseWall}%
\end{table}%
\newpage

\begin{figure}[ht!]
 \includegraphics[width = 8.333cm]{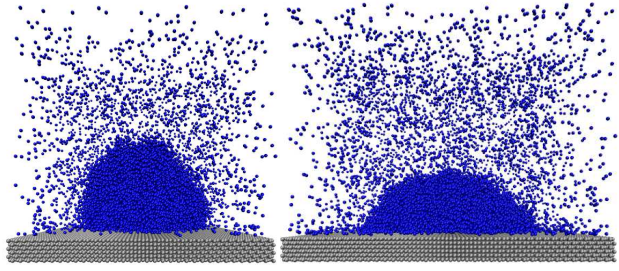}
 \caption{Simulation snapshots taken with VMD\cite{Hump96} at a temperature of $T = 0.8 ~\epsilon_f / k$. The reduced solid--fluid interaction energy
	  is $\zeta = 0.25$ (left) and $\zeta = 0.65$ (right), resulting in different contact angles.}
 \label{fig:Snapshots}
\end{figure}
\newpage

\begin{figure}[htb]
 \includegraphics[width = 8.333cm]{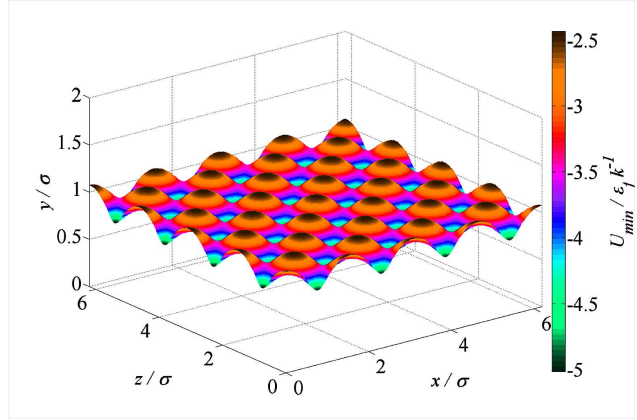}
 \caption{Landscape of the minimal solid--fluid potential, where the normal coordinate $y$ is varied for given values of $x$ and $z$. 
	  The topography of the minimal potential is represented by the 3--dimensional surface. 
	  The color indicates the value of the solid--fluid potential energy. The structure of the lattice can be discerned.}
 \label{fig:nrgLandscapeWall}
\end{figure}
\newpage

\begin{figure}[htb]
 \includegraphics[width = 8.333cm]{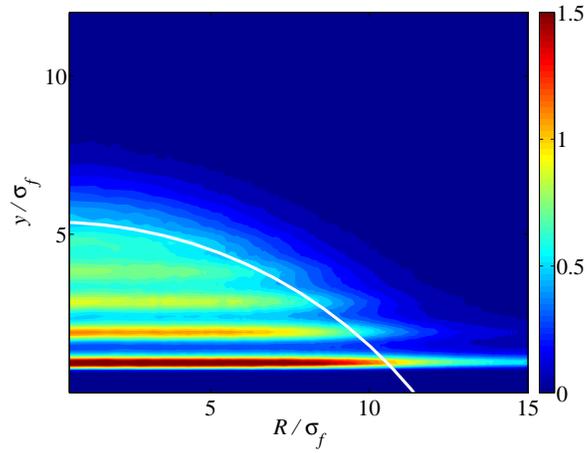}
 \caption{Density profile of a liquid drop with a total number of 1500 fluid particles. 
	  The temperature and the reduced solid--fluid interaction energy are $T = 0.8~\epsilon_f/k$ and $\zeta = 0.65$, respectively. 
	  The small droplet is perturbed by layering effects to which the decrease in the contact angle can be attributed.
	  (The length scale is different from Figure~2.)} 
 \label{fig:SizeEffDensityProfile}
\end{figure}
\newpage


\begin{figure}[htb]
 \includegraphics[width = 8.333cm]{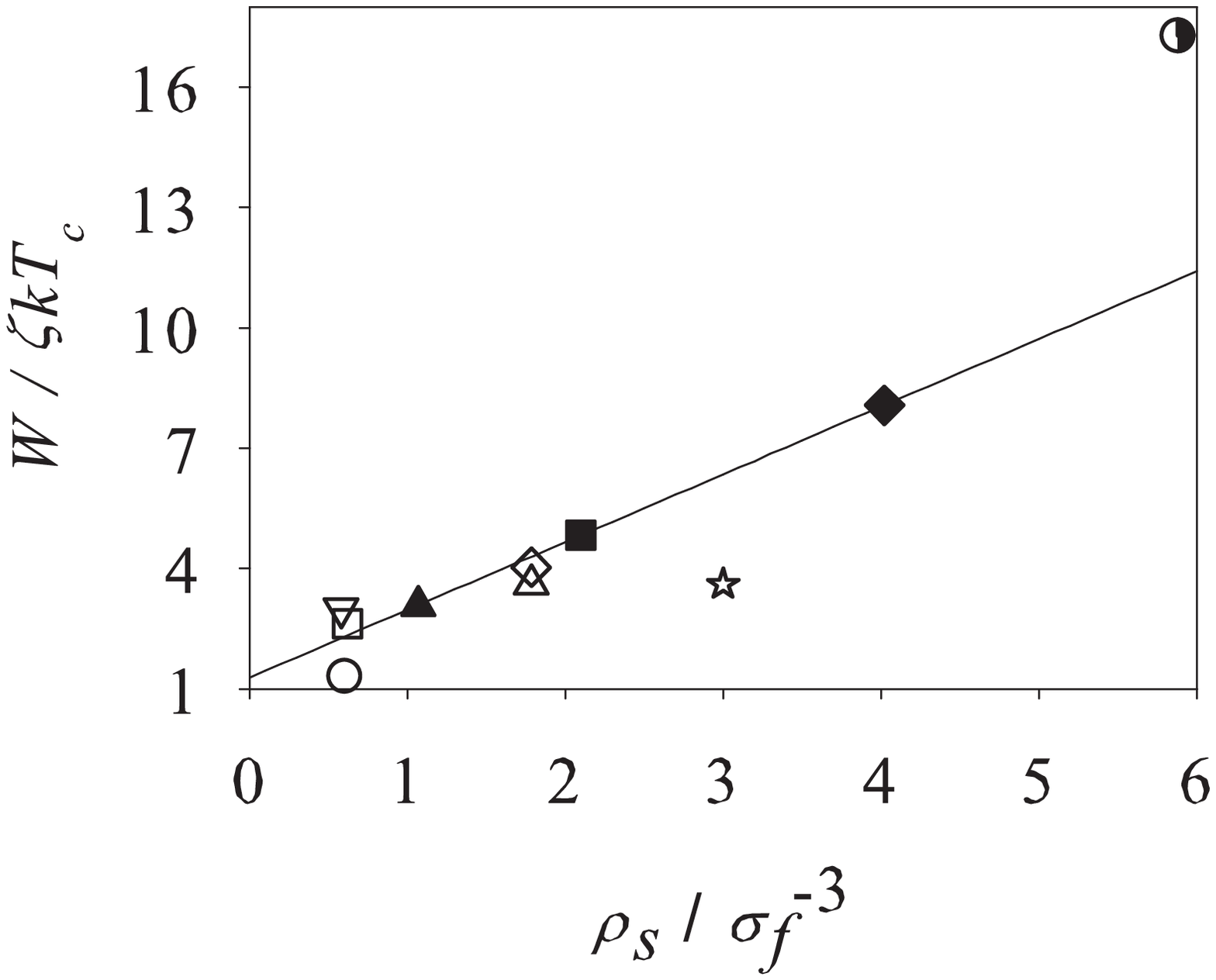}
 \caption{Average minimum potential as a function of the solid density $\rho_s$.
	  This work: $\rho_s = 1.07 \sigma_f^{-3}$ ($\blacktriangle$), 
		     $\rho_s = 2.10 \sigma_f^{-3}$ ($\blacksquare$), 
		     $\rho_s = 4.02 \sigma_f^{-3}$ ($\Diamondblack$). 
	 Ingebrigtsen and Toxvaerd \cite{Ingebrigtsen2007} ($\Circle$);
	 Shaharaz et al. \cite{Shahraz2012} (\ding{73});
	 Grzelak et al. \cite{Grzelak2010a} bcc~(100) lattice ($\triangledown$);
	 Tang and Harris \cite{Tang1995} ($\square$);
	 Nijmeijer et al. \cite{Nijmeijer1990} ($\triangle$);
	 Nijmeijer et al. \cite{Nijmeijer1992} combined model ($\Diamond$);
	 Horsch et al.~\cite{Horsch2010}~($\RIGHTcircle$).
	 The line represents Eq.~(5).}
 \label{fig:WOverRhoS}
\end{figure}
\newpage

\begin{table}[htb]
  \begin{center}
  \caption{Density of the liquid phase $\rho'$ from correlation Eq.~(8).} 
    \begin{tabular}{rrrrrr}
    \toprule
    \multicolumn{1}{c}{$\zeta$} & \multicolumn{1}{c}{$T = 0.7~\epsilon_f / k$} & \multicolumn{1}{c}{$T = 0.8~\epsilon_f / k$} & \multicolumn{1}{c}{$T = 0.9~\epsilon_f / k$} & \multicolumn{1}{c}{$T = 0.95~\epsilon_f / k$} & \multicolumn{1}{c}{$T = 1.0~\epsilon_f / k$} \\
    \midrule
    0.25  & 0.7922 & 0.7379 & 0.6773 & \multicolumn{1}{c}{} &  \\
    0.3   & 0.7901 & 0.7340 & 0.6735 & 0.6364 & 0.5857 \\
    0.35  & 0.7885 & 0.7320 & 0.6699 & 0.6367 & 0.5994 \\
    0.4   & 0.7883 & 0.7315 & 0.6649 & 0.6302 & 0.5917 \\
    0.5   & 0.7892 & 0.7317 & 0.6604 & 0.6250 & 0.5611 \\
    0.6   & 0.7915 & 0.7360 & 0.6693 & 0.6340 & 0.5567 \\
    0.65  & 0.7939 & 0.7383 & 0.6755 & 0.6887 &  \\
    0.75  & 0.8083 & 0.7627 &       & \multicolumn{1}{c}{} &  \\
    \bottomrule
    \end{tabular}%
  \label{tab:fitRhoLiq}%
  \end{center}
\end{table}%
\newpage

\begin{table}[htb]
  \centering
  \caption{Density of the vapor phase $\rho''$  from correlation Eq.~(8).}
    \begin{tabular}{rrrrrr}
    \toprule
    \multicolumn{1}{c}{$\zeta$} & \multicolumn{1}{c}{$T = 0.7~\epsilon_f / k$} & \multicolumn{1}{c}{$T = 0.8~\epsilon_f / k$} & \multicolumn{1}{c}{$T = 0.9~\epsilon_f / k$} & \multicolumn{1}{c}{$T = 0.95~\epsilon_f / k$} & \multicolumn{1}{c}{$T = 1.0~\epsilon_f / k$} \\
    \midrule
    0.25  & 0.00824 & 0.02219 & 0.04835 & \multicolumn{1}{c}{} &  \\
    0.3   & 0.00803 & 0.02235 & 0.05053 & 0.07260 & 0.10481 \\
    0.35  & 0.00834 & 0.02241 & 0.05012 & 0.07425 & 0.10424 \\
    0.4   & 0.00829 & 0.02231 & 0.04992 & 0.07160 & 0.10353 \\
    0.5   & 0.00852 & 0.02230 & 0.04833 & 0.06973 & 0.09742 \\
    0.6   & 0.00915 & 0.02065 & 0.04858 & 0.06559 & 0.09077 \\
    0.65  & 0.00917 & 0.02459 & 0.04521 & 0.06320 &  \\
    0.75  & 0.00932 & 0.01802 &       & \multicolumn{1}{c}{} &  \\
    \bottomrule
    \end{tabular}%
  \label{tab:fitRhoVap}%
\end{table}%
\newpage

\begin{table}[htb]
  \centering
  \caption{Equimolar radius $\mathcal{R}_e$ from correlation Eq.~(8).}
    \begin{tabular}{rrrrrr}
    \toprule
    \multicolumn{1}{c}{$\zeta$} & \multicolumn{1}{c}{$T = 0.7~\epsilon_f / k$} & \multicolumn{1}{c}{$T = 0.8~\epsilon_f / k$} & \multicolumn{1}{c}{$T = 0.9~\epsilon_f / k$} & \multicolumn{1}{c}{$T = 0.95~\epsilon_f / k$} & \multicolumn{1}{c}{$T = 1.0~\epsilon_f / k$} \\
    \midrule
    0.25  & 16.10 & 15.90 & 14.67 & \multicolumn{1}{c}{} &  \\
    0.3   & 16.53 & 16.15 & 14.95 & 13.65 & 12.39 \\
    0.35  & 16.89 & 16.27 & 15.18 & 13.65 & 12.69 \\
    0.4   & 17.40 & 16.85 & 15.65 & 14.24 & 12.25 \\
    0.5   & 19.46 & 18.61 & 18.51 & 15.83 & 15.78 \\
    0.6   & 22.44 & 23.95 & 22.34 & 21.33 & 21.47 \\
    0.65  & 27.16 & 26.62 & 28.66 & 27.64 &  \\
    0.75  & 52.72 & 57.33 &       & \multicolumn{1}{c}{} &  \\
    \bottomrule
    \end{tabular}%
  \label{tab:fitRequi}%
\end{table}%
\newpage

\begin{table}[htb]
  \centering
  \caption{Liquid--vapor interface thickness $D$ from correlation Eq.~(8).}
    \begin{tabular}{rrrrrr}
    \toprule
    \multicolumn{1}{c}{$\zeta$} & \multicolumn{1}{c}{$T = 0.7~\epsilon_f / k$} & \multicolumn{1}{c}{$T = 0.8~\epsilon_f / k$} & \multicolumn{1}{c}{$T = 0.9~\epsilon_f / k$} & \multicolumn{1}{c}{$T = 0.95~\epsilon_f / k$} & \multicolumn{1}{c}{$T = 1.0~\epsilon_f / k$} \\
    \midrule
    0.25  & 2.338 & 3.026 & 5.478 & \multicolumn{1}{c}{} &  \\
    0.3   & 2.318 & 2.966 & 4.163 & 5.629 & 7.822 \\
    0.35  & 2.262 & 2.963 & 4.230 & 4.904 & 6.855 \\
    0.4   & 2.258 & 2.899 & 3.862 & 4.824 & 6.983 \\
    0.5   & 2.253 & 2.858 & 3.910 & 4.613 & 6.149 \\
    0.6   & 2.233 & 2.720 & 3.900 & 4.554 & 7.035 \\
    0.65  & 2.251 & 2.762 & 3.891 & 4.024 &  \\
    0.75  & 2.352 & 2.928 &       & \multicolumn{1}{c}{} &  \\
    \bottomrule
    \end{tabular}%
  \label{tab:fitInterfThickn}%
\end{table}%
\newpage

\begin{table}[htb]
  \centering
  \caption{Oscillation amplitude $A$ from correlation Eq.~(8).}
    \begin{tabular}{rrrrrr}
    \toprule
    \multicolumn{1}{c}{$\zeta$} & \multicolumn{1}{c}{$T = 0.7~\epsilon_f / k$} & \multicolumn{1}{c}{$T = 0.8~\epsilon_f / k$} & \multicolumn{1}{c}{$T = 0.9~\epsilon_f / k$} & \multicolumn{1}{c}{$T = 0.95~\epsilon_f / k$} & \multicolumn{1}{c}{$T = 1.0~\epsilon_f / k$} \\
    \midrule
    0.25  &       &  	   &       & \multicolumn{1}{c}{} &  \\
    0.3   & 1.076 &       &       &       &        \\
    0.35  & 1.025 &       &       &       &       \\
    0.4   & 1.024 & 1.074 & 1.206 & 1.240 & 1.248 \\
    0.5   & 1.138 & 1.086 & 1.085 & 1.090 & 1.143 \\
    0.6   & 1.270 & 1.166 & 1.128 & 1.115 & 1.130 \\
    0.65  & 1.352 & 1.225 & 1.178 & 1.109 &  \\
    0.75  & 1.543 & 1.386 &       & \multicolumn{1}{c}{} &  \\
    \bottomrule
    \end{tabular}%
  \label{tab:fitOscAmpl}%
\end{table}%
\newpage

\begin{table}[htb]
  \centering
  \caption{Period of the density oscillations $p$ from correlation Eq.~(8).}
    \begin{tabular}{rrrrrr}
    \toprule
    \multicolumn{1}{c}{$\zeta$} & \multicolumn{1}{c}{$T = 0.7~\epsilon_f / k$} & \multicolumn{1}{c}{$T = 0.8~\epsilon_f / k$} & \multicolumn{1}{c}{$T = 0.9~\epsilon_f / k$} & \multicolumn{1}{c}{$T = 0.95~\epsilon_f / k$} & \multicolumn{1}{c}{$T = 1.0~\epsilon_f / k$} \\
    \midrule
    0.25  &    	  &   	   &       & \multicolumn{1}{c}{} &  \\
    0.3   & 0.90  &       &       &       &     \\
    0.35  & 0.91  &       &       &       &      \\
    0.4   & 0.92  & 0.90  & 0.89  & 0.91  & 0.90 \\
    0.5   & 0.93  & 0.93  & 0.90  & 0.91  & 0.89 \\
    0.6   & 0.93  & 0.95  & 0.94  & 0.92  & 0.93 \\
    0.65  & 0.93  & 0.94  & 0.94  & 0.93  &  \\
    0.75  & 0.92  & 0.94  &       & \multicolumn{1}{c}{} &  \\
    \bottomrule
    \end{tabular}%
  \label{tab:fitOscPeriod}%
\end{table}%
\newpage

\begin{table}[htb]
  \centering
  \caption{Shift parameter $s$ of the density oscillation from correlation Eq.~(8).}
    \begin{tabular}{rrrrrr}
    \toprule
    \multicolumn{1}{c}{$\zeta$} & \multicolumn{1}{c}{$T = 0.7~\epsilon_f / k$} & \multicolumn{1}{c}{$T = 0.8~\epsilon_f / k$} & \multicolumn{1}{c}{$T = 0.9~\epsilon_f / k$} & \multicolumn{1}{c}{$T = 0.95~\epsilon_f / k$} & \multicolumn{1}{c}{$T = 1.0~\epsilon_f / k$} \\
    \midrule
    0.25  &    	  &   	   &       & \multicolumn{1}{c}{} &  \\
    0.3   & 0.552 &       &       &       &     \\
    0.35  & 0.503 &       &       &       &     \\
    0.4   & 0.446 & 0.527 & 0.605 & 0.621 & 0.622 \\
    0.5   & 0.352 & 0.426 & 0.501 & 0.522 & 0.557 \\
    0.6   & 0.277 & 0.341 & 0.413 & 0.450 & 0.467 \\
    0.65  & 0.242 & 0.311 & 0.369 & 0.415 &  \\
    0.75  & 0.187 & 0.253 &       & \multicolumn{1}{c}{} &  \\
    \bottomrule
    \end{tabular}%
  \label{tab:fitShift}%
\end{table}%
\newpage

\begin{table}[htb]
  \centering
  \caption{Damping parameter $c$ of the density oscillation from correlation Eq.~(8).}
    \begin{tabular}{rrrrrr}
    \toprule
   \multicolumn{1}{c}{$\zeta$} & \multicolumn{1}{c}{$T = 0.7~\epsilon_f / k$} & \multicolumn{1}{c}{$T = 0.8~\epsilon_f / k$} & \multicolumn{1}{c}{$T = 0.9~\epsilon_f / k$} & \multicolumn{1}{c}{$T = 0.95~\epsilon_f / k$} & \multicolumn{1}{c}{$T = 1.0~\epsilon_f / k$} \\
    0.25  &    	  &   	   &       & \multicolumn{1}{c}{} &  \\
    0.3   & 1.570 &       &       &       &     \\
    0.35  & 1.065 &       &       &       &     \\
    0.4   & 0.855 & 1.394 & 2.655 & 3.064 & 3.026 \\
    0.5   & 0.683 & 0.927 & 1.342 & 1.415 & 1.808 \\
    0.6   & 0.629 & 0.760 & 0.962 & 1.088 & 1.208 \\
    0.65  & 0.631 & 0.743 & 0.921 & 1.051 &  \\
    0.75  & 0.579 & 0.720 &       & \multicolumn{1}{c}{} &  \\
    \bottomrule
    \end{tabular}%
  \label{tab:fitDamping}%
\end{table}%
\newpage

\begin{table}[htb]
  \centering
  \caption{Normal coordinate $y(\mathcal{R} =0)$ of the centre of the droplet from correlation Eq.~(8).}
    \begin{tabular}{rrrrrr}
    \toprule
    \multicolumn{1}{c}{$\zeta$} & \multicolumn{1}{c}{$T = 0.7~\epsilon_f / k$} & \multicolumn{1}{c}{$T = 0.8~\epsilon_f / k$} & \multicolumn{1}{c}{$T = 0.9~\epsilon_f / k$} & \multicolumn{1}{c}{$T = 0.95~\epsilon_f / k$} & \multicolumn{1}{c}{$T = 1.0~\epsilon_f / k$} \\
    \midrule
     0.25  & 13.33 & 14.87 & 18.96 &   \multicolumn{1}{c}{} &  \\
    0.3   & 11.99 & 12.39 & 14.39 & 14.54 & 19.25 \\
    0.35  & 9.94  & 11.25 & 12.10 & 13.04 & 12.81 \\
    0.4   & 7.77  & 8.78  & 9.51  & 10.00 & 11.02 \\
    0.5   & 2.21  & 2.76  & 1.43  & 3.98  & 2.13 \\
    0.6   & -4.51 & -7.33 & -7.51 & -8.73 & -13.53 \\
    0.65  & -11.59 & -12.43 & -16.31 & -19.51 &  \\
    0.75  & -42.81 & -48.76 &    & \multicolumn{1}{c}{} &  \\
    \bottomrule
    \end{tabular}%
  \label{tab:CircCentre}%
\end{table}%


\begin{mcitethebibliography}{57}
\providecommand*\natexlab[1]{#1}
\providecommand*\mciteSetBstSublistMode[1]{}
\providecommand*\mciteSetBstMaxWidthForm[2]{}
\providecommand*\mciteBstWouldAddEndPuncttrue
  {\def\EndOfBibitem{\unskip.}}
\providecommand*\mciteBstWouldAddEndPunctfalse
  {\let\EndOfBibitem\relax}
\providecommand*\mciteSetBstMidEndSepPunct[3]{}
\providecommand*\mciteSetBstSublistLabelBeginEnd[3]{}
\providecommand*\EndOfBibitem{}
\mciteSetBstSublistMode{f}
\mciteSetBstMaxWidthForm{subitem}{(\alph{mcitesubitemcount})}
\mciteSetBstSublistLabelBeginEnd
  {\mcitemaxwidthsubitemform\space}
  {\relax}
  {\relax}

\bibitem[Jorgensen and Tirado-Rives(1988)Jorgensen, and
  Tirado-Rives]{Jorgensen1988}
Jorgensen,~W.~L.; Tirado-Rives,~J. The OPLS [optimized potentials for liquid
  simulations] potential functions for proteins, energy minimizations for
  crystals of cyclic peptides and crambin. \emph{J. Am. Chem. Soc.}
  \textbf{1988}, \emph{110}, 1657--1666\relax
\mciteBstWouldAddEndPuncttrue
\mciteSetBstMidEndSepPunct{\mcitedefaultmidpunct}
{\mcitedefaultendpunct}{\mcitedefaultseppunct}\relax
\EndOfBibitem
\bibitem[Jorgensen et~al.(1996)Jorgensen, Maxwell, and
  Tirado-Rives]{Jorgensen1996}
Jorgensen,~W.~L.; Maxwell,~D.~S.; Tirado-Rives,~J. Development and Testing of
  the OPLS All-Atom Force Field on Conformational Energetics and Properties of
  Organic Liquids. \emph{J. Am. Chem. Soc.} \textbf{1996}, \emph{118},
  11225--11236\relax
\mciteBstWouldAddEndPuncttrue
\mciteSetBstMidEndSepPunct{\mcitedefaultmidpunct}
{\mcitedefaultendpunct}{\mcitedefaultseppunct}\relax
\EndOfBibitem
\bibitem[Martin and Siepmann(1998)Martin, and Siepmann]{Martin1998}
Martin,~M.~G.; Siepmann,~J.~I. Transferable Potentials for Phase Equilibria. 1.
  United-Atom Description of n-Alkanes. \emph{J. Phys. Chem. B} \textbf{1998},
  \emph{102}, 2569--2577\relax
\mciteBstWouldAddEndPuncttrue
\mciteSetBstMidEndSepPunct{\mcitedefaultmidpunct}
{\mcitedefaultendpunct}{\mcitedefaultseppunct}\relax
\EndOfBibitem
\bibitem[Keasler et~al.(2012)Keasler, Charan, Wick, Economou, and
  Siepmann]{Keasler2012}
Keasler,~S.~J.; Charan,~S.~M.; Wick,~C.~D.; Economou,~I.~G.; Siepmann,~J.~I.
  Transferable Potentials for Phase Equilibria--United Atom Description of
  Five- and Six-Membered Cyclic Alkanes and Ethers. \emph{J. Phys. Chem. B}
  \textbf{2012}, \emph{116}, 11234--11246, PMID: 22900670\relax
\mciteBstWouldAddEndPuncttrue
\mciteSetBstMidEndSepPunct{\mcitedefaultmidpunct}
{\mcitedefaultendpunct}{\mcitedefaultseppunct}\relax
\EndOfBibitem
\bibitem[Kiyohara et~al.(1998)Kiyohara, Gubbins, and
  Panagiotopoulos]{PANAGIOTOPOULOS1998}
Kiyohara,~K.; Gubbins,~K.; Panagiotopoulos,~A. Phase coexistence properties of
  polarizable water models. \emph{Mol. Phys.} \textbf{1998}, \emph{94},
  803--808\relax
\mciteBstWouldAddEndPuncttrue
\mciteSetBstMidEndSepPunct{\mcitedefaultmidpunct}
{\mcitedefaultendpunct}{\mcitedefaultseppunct}\relax
\EndOfBibitem
\bibitem[Potoff et~al.(1999)Potoff, Errington, and Panagiotopoulos]{POTOFF1999}
Potoff,~J.~J.; Errington,~J.~R.; Panagiotopoulos,~A. Molecular simulation of
  phase equilibria for mixtures of polar and non-polar components. \emph{Mol.
  Phys.} \textbf{1999}, \emph{97}, 1073--1083\relax
\mciteBstWouldAddEndPuncttrue
\mciteSetBstMidEndSepPunct{\mcitedefaultmidpunct}
{\mcitedefaultendpunct}{\mcitedefaultseppunct}\relax
\EndOfBibitem
\bibitem[MacKerell et~al.(2000)MacKerell, Banavali, and Foloppe]{MacKerell2000}
MacKerell,~A.~D.; Banavali,~N.; Foloppe,~N. Development and current status of
  the CHARMM force field for nucleic acids. \emph{Biopolymers} \textbf{2000},
  \emph{56}, 257--265\relax
\mciteBstWouldAddEndPuncttrue
\mciteSetBstMidEndSepPunct{\mcitedefaultmidpunct}
{\mcitedefaultendpunct}{\mcitedefaultseppunct}\relax
\EndOfBibitem
\bibitem[Vrabec et~al.(2001)Vrabec, Stoll, and Hasse]{Vrabec2001}
Vrabec,~J.; Stoll,~J.; Hasse,~H. A Set of Molecular Models for Symmetric
  Quadrupolar Fluids. \emph{J. Phys. Chem. B} \textbf{2001}, \emph{105},
  12126--12133\relax
\mciteBstWouldAddEndPuncttrue
\mciteSetBstMidEndSepPunct{\mcitedefaultmidpunct}
{\mcitedefaultendpunct}{\mcitedefaultseppunct}\relax
\EndOfBibitem
\bibitem[Stoll et~al.(2003)Stoll, Vrabec, and Hasse]{Stoll2003}
Stoll,~J.; Vrabec,~J.; Hasse,~H. Comprehensive study of the vapour-liquid
  equilibria of the pure two-centre Lennard-Jones plus pointdipole fluid.
  \emph{Fluid Phase Equilib.} \textbf{2003}, \emph{209}, 29--53\relax
\mciteBstWouldAddEndPuncttrue
\mciteSetBstMidEndSepPunct{\mcitedefaultmidpunct}
{\mcitedefaultendpunct}{\mcitedefaultseppunct}\relax
\EndOfBibitem
\bibitem[Vorholz et~al.(2004)Vorholz, Harismiadis, Panagiotopoulos, Rumpf, and
  Maurer]{Vorholz2004}
Vorholz,~J.; Harismiadis,~V.; Panagiotopoulos,~A.; Rumpf,~B.; Maurer,~G.
  Molecular simulation of the solubility of carbon dioxide in aqueous solutions
  of sodium chloride. \emph{Fluid Phase Equilib.} \textbf{2004}, \emph{226},
  237--250\relax
\mciteBstWouldAddEndPuncttrue
\mciteSetBstMidEndSepPunct{\mcitedefaultmidpunct}
{\mcitedefaultendpunct}{\mcitedefaultseppunct}\relax
\EndOfBibitem
\bibitem[Moghaddam and Panagiotopoulos(2004)Moghaddam, and
  Panagiotopoulos]{Moghaddam2004}
Moghaddam,~S.; Panagiotopoulos,~A.~Z. Determination of second virial
  coefficients by grand canonical Monte Carlo simulations. \emph{Fluid Phase
  Equilib.} \textbf{2004}, \emph{222--223}, 221--224\relax
\mciteBstWouldAddEndPuncttrue
\mciteSetBstMidEndSepPunct{\mcitedefaultmidpunct}
{\mcitedefaultendpunct}{\mcitedefaultseppunct}\relax
\EndOfBibitem
\bibitem[Deublein et~al.(2012)Deublein, Vrabec, and Hasse]{Deublein2012}
Deublein,~S.; Vrabec,~J.; Hasse,~H. A set of molecular models for alkali and
  halide ions in aqueous solution. \emph{J. Chem. Phys.} \textbf{2012},
  \emph{136}, 084501\relax
\mciteBstWouldAddEndPuncttrue
\mciteSetBstMidEndSepPunct{\mcitedefaultmidpunct}
{\mcitedefaultendpunct}{\mcitedefaultseppunct}\relax
\EndOfBibitem
\bibitem[Merker et~al.(2012)Merker, Vrabec, and Hasse]{Merker2012}
Merker,~T.; Vrabec,~J.; Hasse,~H. Engineering Molecular Models: Efficient
  Parameterization Procedure and Cyclohexanol as Case Study. \emph{Soft Mater.}
  \textbf{2012}, \emph{10}, 3--25\relax
\mciteBstWouldAddEndPuncttrue
\mciteSetBstMidEndSepPunct{\mcitedefaultmidpunct}
{\mcitedefaultendpunct}{\mcitedefaultseppunct}\relax
\EndOfBibitem
\bibitem[Schapotschnikow et~al.(2007)Schapotschnikow, Pool, and
  Vlugt]{Schapotschnikow2007a}
Schapotschnikow,~P.; Pool,~R.; Vlugt,~T. J.~H. Selective adsorption of alkyl
  thiols on gold in different geometries. \emph{Comput. Phys. Commun.}
  \textbf{2007}, \emph{177}, 154--157\relax
\mciteBstWouldAddEndPuncttrue
\mciteSetBstMidEndSepPunct{\mcitedefaultmidpunct}
{\mcitedefaultendpunct}{\mcitedefaultseppunct}\relax
\EndOfBibitem
\bibitem[Steele(1974)]{Steele1974}
Steele,~W.~A. \emph{The Interaction of Gases with Solid Surfaces}, 1st ed.;
  Pergamon: Oxford, 1974\relax
\mciteBstWouldAddEndPuncttrue
\mciteSetBstMidEndSepPunct{\mcitedefaultmidpunct}
{\mcitedefaultendpunct}{\mcitedefaultseppunct}\relax
\EndOfBibitem
\bibitem[Findenegg and Fischer(1975)Findenegg, and Fischer]{Findenegg1975}
Findenegg,~G.~H.; Fischer,~J. Adsorption of fluids: simple theories for the
  density profile in a fluid near an adsorbing surface. \emph{Faraday Discuss.
  Chem. Soc.} \textbf{1975}, \emph{59}, 38--45\relax
\mciteBstWouldAddEndPuncttrue
\mciteSetBstMidEndSepPunct{\mcitedefaultmidpunct}
{\mcitedefaultendpunct}{\mcitedefaultseppunct}\relax
\EndOfBibitem
\bibitem[Fischer et~al.(1982)Fischer, Bohn, Körner, and Findenegg]{Fischer1982}
Fischer,~J.; Bohn,~M.; Körner,~B.; Findenegg,~G.~H. Gasadsorption in Poren.
  \emph{Chem. Ing. Tech.} \textbf{1982}, \emph{54}, 763--763\relax
\mciteBstWouldAddEndPuncttrue
\mciteSetBstMidEndSepPunct{\mcitedefaultmidpunct}
{\mcitedefaultendpunct}{\mcitedefaultseppunct}\relax
\EndOfBibitem
\bibitem[Bucior et~al.(2009)Bucior, Yelash, and Binder]{Bucior2009}
Bucior,~K.; Yelash,~L.; Binder,~K. Molecular-dynamics simulation of evaporation
  processes of fluid bridges confined in slitlike pores. \emph{Phys. Rev. E}
  \textbf{2009}, \emph{79}, 031604\relax
\mciteBstWouldAddEndPuncttrue
\mciteSetBstMidEndSepPunct{\mcitedefaultmidpunct}
{\mcitedefaultendpunct}{\mcitedefaultseppunct}\relax
\EndOfBibitem
\bibitem[Schapotschnikow et~al.(2009)Schapotschnikow, Hommersom, and
  Vlugt]{Schapotschnikow2009}
Schapotschnikow,~P.; Hommersom,~B.; Vlugt,~T. J.~H. Adsorption and Binding of
  Ligands to CdSe Nanocrystals. \emph{J. Phys. Chem. C} \textbf{2009},
  \emph{113}, 12690--12698\relax
\mciteBstWouldAddEndPuncttrue
\mciteSetBstMidEndSepPunct{\mcitedefaultmidpunct}
{\mcitedefaultendpunct}{\mcitedefaultseppunct}\relax
\EndOfBibitem
\bibitem[Sokolowski and Fischer(1990)Sokolowski, and Fischer]{Sokolowski1990}
Sokolowski,~S.; Fischer,~J. Classical multicomponent fluid structure near solid
  substrates: Born-Green-Yvon equation versus density-functional theory.
  \emph{Mol. Phys.} \textbf{1990}, \emph{70}, 1097--1113\relax
\mciteBstWouldAddEndPuncttrue
\mciteSetBstMidEndSepPunct{\mcitedefaultmidpunct}
{\mcitedefaultendpunct}{\mcitedefaultseppunct}\relax
\EndOfBibitem
\bibitem[Sikkenk et~al.(1987)Sikkenk, Indekeu, van Leeuwen, and
  Vossnack]{Sikkenk1987}
Sikkenk,~J.~H.; Indekeu,~J.~O.; van Leeuwen,~J. M.~J.; Vossnack,~E.~O.
  Molecular-dynamics simulation of wetting and drying at solid-fluid
  interfaces. \emph{Phys. Rev. Lett.} \textbf{1987}, \emph{59}, 98--101\relax
\mciteBstWouldAddEndPuncttrue
\mciteSetBstMidEndSepPunct{\mcitedefaultmidpunct}
{\mcitedefaultendpunct}{\mcitedefaultseppunct}\relax
\EndOfBibitem
\bibitem[Sikkenk et~al.(1988)Sikkenk, Indekeu, van Leeuwen, Vossnack, and
  Bakker]{Sikkenk1988}
Sikkenk,~J.; Indekeu,~J.; van Leeuwen,~J.; Vossnack,~E.; Bakker,~A. Simulation
  of wetting and drying at solid-fluid interfaces on the Delft Molecular
  Dynamics Processor. \emph{J. Stat. Phys.} \textbf{1988}, \emph{52},
  23--44\relax
\mciteBstWouldAddEndPuncttrue
\mciteSetBstMidEndSepPunct{\mcitedefaultmidpunct}
{\mcitedefaultendpunct}{\mcitedefaultseppunct}\relax
\EndOfBibitem
\bibitem[Nijmeijer et~al.(1989)Nijmeijer, Bruin, Bakker, and van
  Leeuwen]{Nijmeijer1989}
Nijmeijer,~M.; Bruin,~C.; Bakker,~A.; van Leeuwen,~J. A visual measurement of
  contact angles in a molecular-dynamics simulation. \emph{Physica A}
  \textbf{1989}, \emph{160}, 166--180\relax
\mciteBstWouldAddEndPuncttrue
\mciteSetBstMidEndSepPunct{\mcitedefaultmidpunct}
{\mcitedefaultendpunct}{\mcitedefaultseppunct}\relax
\EndOfBibitem
\bibitem[Nijmeijer et~al.(1990)Nijmeijer, Bruin, Bakker, and van
  Leeuwen]{Nijmeijer1990}
Nijmeijer,~M. J.~P.; Bruin,~C.; Bakker,~A.~F.; van Leeuwen,~J. M.~J. Wetting
  and drying of an inert wall by a fluid in a molecular-dynamics simulation.
  \emph{Phys. Rev. A} \textbf{1990}, \emph{42}, 6052--6059\relax
\mciteBstWouldAddEndPuncttrue
\mciteSetBstMidEndSepPunct{\mcitedefaultmidpunct}
{\mcitedefaultendpunct}{\mcitedefaultseppunct}\relax
\EndOfBibitem
\bibitem[Nijmeijer et~al.(1992)Nijmeijer, Bruin, Bakker, and van
  Leeuwen]{Nijmeijer1992}
Nijmeijer,~M.; Bruin,~C.; Bakker,~A.; van Leeuwen,~J. Molecular dynamics of the
  wetting and drying of a wall with a long-ranged wall-fluid interaction.
  \emph{J. Phys.: Condens. Matter} \textbf{1992}, \emph{4}, 15--31\relax
\mciteBstWouldAddEndPuncttrue
\mciteSetBstMidEndSepPunct{\mcitedefaultmidpunct}
{\mcitedefaultendpunct}{\mcitedefaultseppunct}\relax
\EndOfBibitem
\bibitem[Tang and Harris(1995)Tang, and Harris]{Tang1995}
Tang,~J.~Z.; Harris,~J.~G. Fluid wetting on molecularly rough surfaces.
  \emph{J. Chem. Phys.} \textbf{1995}, \emph{103}, 8201--8208\relax
\mciteBstWouldAddEndPuncttrue
\mciteSetBstMidEndSepPunct{\mcitedefaultmidpunct}
{\mcitedefaultendpunct}{\mcitedefaultseppunct}\relax
\EndOfBibitem
\bibitem[Blake et~al.(1997)Blake, Clarke, De~Coninck, and
  de~Ruijter]{Blake1997}
Blake,~T.~D.; Clarke,~A.; De~Coninck,~J.; de~Ruijter,~M.~J. Contact Angle
  Relaxation during Droplet Spreading: Comparison between Molecular Kinetic
  Theory and Molecular Dynamics. \emph{Langmuir} \textbf{1997}, \emph{13},
  2164--2166\relax
\mciteBstWouldAddEndPuncttrue
\mciteSetBstMidEndSepPunct{\mcitedefaultmidpunct}
{\mcitedefaultendpunct}{\mcitedefaultseppunct}\relax
\EndOfBibitem
\bibitem[Werder et~al.(2001)Werder, Walther, Jaffe, Halicioglu, Noca, and
  Koumoutsakos]{Werder2001}
Werder,~T.; Walther,~J.~H.; Jaffe,~R.~L.; Halicioglu,~T.; Noca,~F.;
  Koumoutsakos,~P. Molecular Dynamics Simulation of Contact Angles of Water
  Droplets in Carbon Nanotubes. \emph{Nano Lett.} \textbf{2001}, \emph{1},
  697--702\relax
\mciteBstWouldAddEndPuncttrue
\mciteSetBstMidEndSepPunct{\mcitedefaultmidpunct}
{\mcitedefaultendpunct}{\mcitedefaultseppunct}\relax
\EndOfBibitem
\bibitem[Werder et~al.(2003)Werder, Walther, Jaffe, Halicioglu, and
  Koumoutsakos]{Werder2003}
Werder,~T.; Walther,~J.~H.; Jaffe,~R.~L.; Halicioglu,~T.; Koumoutsakos,~P. On
  the Water-Carbon Interaction for Use in Molecular Dynamics Simulations of
  Graphite and Carbon Nanotubes. \emph{J. Phys. Chem. B} \textbf{2003},
  \emph{107}, 1345--1352\relax
\mciteBstWouldAddEndPuncttrue
\mciteSetBstMidEndSepPunct{\mcitedefaultmidpunct}
{\mcitedefaultendpunct}{\mcitedefaultseppunct}\relax
\EndOfBibitem
\bibitem[Ingebrigtsen and Toxvaerd(2007)Ingebrigtsen, and
  Toxvaerd]{Ingebrigtsen2007}
Ingebrigtsen,~T.; Toxvaerd,~S. Contact Angles of Lennard-Jones Liquids and
  Droplets on Planar Surfaces. \emph{J. Phys. Chem. C} \textbf{2007},
  \emph{111}, 8518--8523\relax
\mciteBstWouldAddEndPuncttrue
\mciteSetBstMidEndSepPunct{\mcitedefaultmidpunct}
{\mcitedefaultendpunct}{\mcitedefaultseppunct}\relax
\EndOfBibitem
\bibitem[Grzelak et~al.(2010)Grzelak, Shen, and Errington]{Grzelak2010a}
Grzelak,~E.~M.; Shen,~V.~K.; Errington,~J.~R. Molecular Simulation Study of
  Anisotropic Wetting. \emph{Langmuir} \textbf{2010}, \emph{26},
  8274--8281\relax
\mciteBstWouldAddEndPuncttrue
\mciteSetBstMidEndSepPunct{\mcitedefaultmidpunct}
{\mcitedefaultendpunct}{\mcitedefaultseppunct}\relax
\EndOfBibitem
\bibitem[Leroy and M\"uller-Plathe(2010)Leroy, and M\"uller-Plathe]{Leroy2010}
Leroy,~F.; M\"uller-Plathe,~F. Solid-liquid surface free energy of Lennard-Jones
  liquid on smooth and rough surfaces computed by molecular dynamics using the
  phantom-wall method. \emph{J. Chem. Phys.} \textbf{2010}, \emph{133},
  044110\relax
\mciteBstWouldAddEndPuncttrue
\mciteSetBstMidEndSepPunct{\mcitedefaultmidpunct}
{\mcitedefaultendpunct}{\mcitedefaultseppunct}\relax
\EndOfBibitem
\bibitem[Rane et~al.(2011)Rane, Kumar, and Errington]{Rane2011}
Rane,~K.~S.; Kumar,~V.; Errington,~J.~R. Monte Carlo simulation methods for
  computing the wetting and drying properties of model systems. \emph{J. Chem.
  Phys.} \textbf{2011}, \emph{135}, 234102\relax
\mciteBstWouldAddEndPuncttrue
\mciteSetBstMidEndSepPunct{\mcitedefaultmidpunct}
{\mcitedefaultendpunct}{\mcitedefaultseppunct}\relax
\EndOfBibitem
\bibitem[Weijs et~al.(2011)Weijs, Marchand, Andreotti, Lohse, and
  Snoeijer]{Weijs2011}
Weijs,~J.~H.; Marchand,~A.; Andreotti,~B.; Lohse,~D.; Snoeijer,~J.~H. Origin of
  line tension for a Lennard-Jones nanodroplet. \emph{Phys. Fluids}
  \textbf{2011}, \emph{23}, 022001\relax
\mciteBstWouldAddEndPuncttrue
\mciteSetBstMidEndSepPunct{\mcitedefaultmidpunct}
{\mcitedefaultendpunct}{\mcitedefaultseppunct}\relax
\EndOfBibitem
\bibitem[Shahraz et~al.(2012)Shahraz, Borhan, and Fichthorn]{Shahraz2012}
Shahraz,~A.; Borhan,~A.; Fichthorn,~K.~A. A Theory for the Morphological
  Dependence of Wetting on a Physically Patterned Solid Surface.
  \emph{Langmuir} \textbf{2012}, \emph{28}, 14227--14237\relax
\mciteBstWouldAddEndPuncttrue
\mciteSetBstMidEndSepPunct{\mcitedefaultmidpunct}
{\mcitedefaultendpunct}{\mcitedefaultseppunct}\relax
\EndOfBibitem
\bibitem[Allen and Tildesley(2009)Allen, and Tildesley]{Allen2009}
Allen,~M.; Tildesley,~D. \emph{Computer Simulation of Liquids}; Clarendon:
  Oxford, 2009\relax
\mciteBstWouldAddEndPuncttrue
\mciteSetBstMidEndSepPunct{\mcitedefaultmidpunct}
{\mcitedefaultendpunct}{\mcitedefaultseppunct}\relax
\EndOfBibitem
\bibitem[Horsch et~al.(2010)Horsch, Heitzig, Dan, Harting, Hasse, and
  Vrabec]{Horsch2010}
Horsch,~M.; Heitzig,~M.; Dan,~C.; Harting,~J.; Hasse,~H.; Vrabec,~J. Contact
  Angle Dependence on the Fluid--Wall Dispersive Energy. \emph{Langmuir}
  \textbf{2010}, \emph{26}, 10913--10917\relax
\mciteBstWouldAddEndPuncttrue
\mciteSetBstMidEndSepPunct{\mcitedefaultmidpunct}
{\mcitedefaultendpunct}{\mcitedefaultseppunct}\relax
\EndOfBibitem
\bibitem[Vrabec et~al.(2006)Vrabec, Kedia, Fuchs, and Hasse]{Vrabec2006}
Vrabec,~J.; Kedia,~G.~K.; Fuchs,~G.; Hasse,~H. Comprehensive study of the
  vapour-liquid coexistence of the truncated and shifted Lennard--Jones fluid
  including planar and spherical interface properties. \emph{Mol. Phys.}
  \textbf{2006}, \emph{104}, 1509--1527\relax
\mciteBstWouldAddEndPuncttrue
\mciteSetBstMidEndSepPunct{\mcitedefaultmidpunct}
{\mcitedefaultendpunct}{\mcitedefaultseppunct}\relax
\EndOfBibitem
\bibitem[van Meel et~al.(2008)van Meel, Page, Sear, and Frenkel]{Meel2008}
van Meel,~J.~A.; Page,~A.~J.; Sear,~R.~P.; Frenkel,~D. Two-step vapor-crystal
  nucleation close below triple point. \emph{J. Chem. Phys.} \textbf{2008},
  \emph{129}, 204505\relax
\mciteBstWouldAddEndPuncttrue
\mciteSetBstMidEndSepPunct{\mcitedefaultmidpunct}
{\mcitedefaultendpunct}{\mcitedefaultseppunct}\relax
\EndOfBibitem
\bibitem[Hamaker(1937)]{Hamaker1937}
Hamaker,~H. The London--van der Waals attraction between spherical particles.
  \emph{Physica} \textbf{1937}, \emph{4}, 1058 -- 1072\relax
\mciteBstWouldAddEndPuncttrue
\mciteSetBstMidEndSepPunct{\mcitedefaultmidpunct}
{\mcitedefaultendpunct}{\mcitedefaultseppunct}\relax
\EndOfBibitem
\bibitem[Pethica(1977)]{Pethica1977}
Pethica,~B. The contact angle equilibrium. \emph{J. Colloid Interface Sci.}
  \textbf{1977}, \emph{62}, 567 -- 569\relax
\mciteBstWouldAddEndPuncttrue
\mciteSetBstMidEndSepPunct{\mcitedefaultmidpunct}
{\mcitedefaultendpunct}{\mcitedefaultseppunct}\relax
\EndOfBibitem
\bibitem[Tolman(1949)]{Tolman1949}
Tolman,~R.~C. The Effect of Droplet Size on Surface Tension. \emph{J. Chem.
  Phys.} \textbf{1949}, \emph{17}, 333--337\relax
\mciteBstWouldAddEndPuncttrue
\mciteSetBstMidEndSepPunct{\mcitedefaultmidpunct}
{\mcitedefaultendpunct}{\mcitedefaultseppunct}\relax
\EndOfBibitem
\bibitem[Engin et~al.(2008)Engin, Sandoval, and Urbassek]{Engin2008}
Engin,~C.; Sandoval,~L.; Urbassek,~H.~M. Characterization of Fe potentials with
  respect to the stability of the bcc and fcc phase. \emph{Modell. Simul.
  Mater. Sci. Eng.} \textbf{2008}, \emph{16}, 035005\relax
\mciteBstWouldAddEndPuncttrue
\mciteSetBstMidEndSepPunct{\mcitedefaultmidpunct}
{\mcitedefaultendpunct}{\mcitedefaultseppunct}\relax
\EndOfBibitem
\bibitem[Buchholz et~al.(2011)Buchholz, Bungartz, and Vrabec]{Buchholz2011124}
Buchholz,~M.; Bungartz,~H.-J.; Vrabec,~J. Software design for a highly parallel
  molecular dynamics simulation framework in chemical engineering. \emph{J.
  Comput. Sci.} \textbf{2011}, \emph{2}, 124 -- 129\relax
\mciteBstWouldAddEndPuncttrue
\mciteSetBstMidEndSepPunct{\mcitedefaultmidpunct}
{\mcitedefaultendpunct}{\mcitedefaultseppunct}\relax
\EndOfBibitem
\bibitem[Oleinikova et~al.(2006)Oleinikova, Brovchenko, and
  Geiger]{Oleinikova2006}
Oleinikova,~A.; Brovchenko,~I.; Geiger,~A. Behavior of a wetting phase near a
  solid boundary: vapor near a weakly attractive surface. \emph{Eur. Phys. J.
  B} \textbf{2006}, \emph{52}, 507--519\relax
\mciteBstWouldAddEndPuncttrue
\mciteSetBstMidEndSepPunct{\mcitedefaultmidpunct}
{\mcitedefaultendpunct}{\mcitedefaultseppunct}\relax
\EndOfBibitem
\bibitem[Forte et~al.(2014)Forte, Haslam, Jackson, and M{\"u}ller]{Forte2014}
Forte,~E.; Haslam,~A.~J.; Jackson,~G.; M{\"u}ller,~E.~A. Effective
  coarse-grained solid-fluid potentials and their application to model
  adsorption of fluids on heterogeneous surfaces. \emph{Phys. Chem. Chem.
  Phys.} \textbf{2014}, \emph{16}, 19165--19180\relax
\mciteBstWouldAddEndPuncttrue
\mciteSetBstMidEndSepPunct{\mcitedefaultmidpunct}
{\mcitedefaultendpunct}{\mcitedefaultseppunct}\relax
\EndOfBibitem
\bibitem[de~Gennes(1985)]{deGennes85}
de~Gennes,~P.~G. Wetting: statics and dynamics. \emph{Rev. Mod. Phys.}
  \textbf{1985}, \emph{57}, 827--863\relax
\mciteBstWouldAddEndPuncttrue
\mciteSetBstMidEndSepPunct{\mcitedefaultmidpunct}
{\mcitedefaultendpunct}{\mcitedefaultseppunct}\relax
\EndOfBibitem
\bibitem[Werth et~al.(2013)Werth, Lishchuk, Horsch, and Hasse]{Werth2013}
Werth,~S.; Lishchuk,~S.~V.; Horsch,~M.; Hasse,~H. The influence of the liquid
  slab thickness on the planar vapor--liquid interfacial tension. \emph{Physica
  A: Statistical Mechanics and its Applications} \textbf{2013}, \emph{392},
  2359 -- 2367\relax
\mciteBstWouldAddEndPuncttrue
\mciteSetBstMidEndSepPunct{\mcitedefaultmidpunct}
{\mcitedefaultendpunct}{\mcitedefaultseppunct}\relax
\EndOfBibitem
\bibitem[Santiso et~al.(2013)Santiso, Herdes, and M{\"u}ller]{Santiso2013}
Santiso,~E.~E.; Herdes,~C.; M{\"u}ller,~E.~A. On the Calculation of Solid-Fluid
  Contact Angles from Molecular Dynamics. \emph{Entropy} \textbf{2013},
  \emph{15}, 3734--3745\relax
\mciteBstWouldAddEndPuncttrue
\mciteSetBstMidEndSepPunct{\mcitedefaultmidpunct}
{\mcitedefaultendpunct}{\mcitedefaultseppunct}\relax
\EndOfBibitem
\bibitem[Rowlinson and Widom(2002)Rowlinson, and Widom]{Rowlinson02}
Rowlinson,~J.; Widom,~B. \emph{Molecular Theory of Capillarity}; Dover: New
  York, 2002\relax
\mciteBstWouldAddEndPuncttrue
\mciteSetBstMidEndSepPunct{\mcitedefaultmidpunct}
{\mcitedefaultendpunct}{\mcitedefaultseppunct}\relax
\EndOfBibitem
\bibitem[Brovchenko and Oleinikova(2012)Brovchenko, and
  Oleinikova]{Brovchenko12}
Brovchenko,~I.; Oleinikova,~A. Universal Shape of the Fluid Density Profiles
  Near a Solid Boundary: LJ Vapor Near Weakly Attractive and Hard Walls.
  \emph{Soft Mater.} \textbf{2012}, \emph{10}, 106--129\relax
\mciteBstWouldAddEndPuncttrue
\mciteSetBstMidEndSepPunct{\mcitedefaultmidpunct}
{\mcitedefaultendpunct}{\mcitedefaultseppunct}\relax
\EndOfBibitem
\bibitem[Henderson and van Swol(1990)Henderson, and van Swol]{Henderson1990}
Henderson,~J.~R.; van Swol,~F. Fluctuation phenomena at a first--order phase
  transition. \emph{Journal of Physics: Condensed Matter} \textbf{1990},
  \emph{2}, 4537\relax
\mciteBstWouldAddEndPuncttrue
\mciteSetBstMidEndSepPunct{\mcitedefaultmidpunct}
{\mcitedefaultendpunct}{\mcitedefaultseppunct}\relax
\EndOfBibitem
\bibitem[Monson(2008)]{Monson2008}
Monson,~P.~A. Contact Angles, Pore Condensation, and Hysteresis: Insights from
  a Simple Molecular Model. \emph{Langmuir} \textbf{2008}, \emph{24},
  12295--12302\relax
\mciteBstWouldAddEndPuncttrue
\mciteSetBstMidEndSepPunct{\mcitedefaultmidpunct}
{\mcitedefaultendpunct}{\mcitedefaultseppunct}\relax
\EndOfBibitem
\bibitem[Young(1805)]{Young1805}
Young,~T. An Essay on the Cohesion of Fluids. \emph{Phil. Trans. R. Soc. Lond.}
  \textbf{1805}, \emph{95}, 65--87\relax
\mciteBstWouldAddEndPuncttrue
\mciteSetBstMidEndSepPunct{\mcitedefaultmidpunct}
{\mcitedefaultendpunct}{\mcitedefaultseppunct}\relax
\EndOfBibitem
\bibitem[Horsch et~al.(2010)Horsch, Miroshnichenko, Vrabec, Glass, Niethammer, Bernreuther, M{\"u}ller, and Jackson]{Horsch2010a}
Horsch,~M.~T.; Miroshnichenko,~S.; Vrabec,~J.; Glass,~C.; Niethammer,~C.;
  Bernreuther,~M.; M{\"u}ller,~E.~A.; Jackson,~G. In \emph{Competence in High
  Performance Computing 2010}; Bischof,~C., Hegering,~H.-G., Nagel,~W.,
  Wittum,~G., Eds.; Springer, 2010; pp 73--84\relax
\mciteBstWouldAddEndPuncttrue
\mciteSetBstMidEndSepPunct{\mcitedefaultmidpunct}
{\mcitedefaultendpunct}{\mcitedefaultseppunct}\relax
\EndOfBibitem
\bibitem[Humphrey et~al.(1996)Humphrey, Dalke, and Schulten]{Hump96}
Humphrey,~W.; Dalke,~A.; Schulten,~K. {VMD} -- {V}isual {M}olecular {D}ynamics.
  \emph{Journal of Molecular Graphics} \textbf{1996}, \emph{14}, 33--38\relax
\mciteBstWouldAddEndPuncttrue
\mciteSetBstMidEndSepPunct{\mcitedefaultmidpunct}
{\mcitedefaultendpunct}{\mcitedefaultseppunct}\relax
\EndOfBibitem
\end{mcitethebibliography}
\end{document}